\documentclass[sn-basic]{sn-jnl}% Basic Springer Nature Reference Style/Chemistry Reference Style

\jyear{2022}%
\usepackage{amssymb,latexsym,amsmath,extarrows, mathrsfs, amsthm}
\usepackage{amssymb}
\usepackage{hyperref}
\usepackage{epstopdf}
\usepackage[utf8]{inputenc}
\theoremstyle{thmstyleone}%
\newtheorem{theorem}{Theorem}%  meant for continuous numbers
\newtheorem{assumption}{Assumption}[section]

\newtheorem{corollary}{Corollary}

\theoremstyle{thmstyletwo}%

\theoremstyle{thmstylethree}%

\begin{document}

\title[Single Parameter Inference of  Non-sparse Logistic Regression Models]{Single Parameter Inference of  Non-sparse Logistic Regression Models}


\author[1]{\fnm{Yanmei} \sur{Shi}}\email{2020020263@qdu.edu.cn}
\author*[1]{\fnm{Qi} \sur{Zhang}}\email{qizhang@qdu.edu.cn}


%\affil[1]{School of Mathematics and Statistics, Qingdao University, 308 Ningxia Road, Shinan District, Qingdao, Shandong, China}

\affil[1]{Institute of Mathematics and Statistics, Qingdao University, 308 Ningxia Road, Shinan District, Qingdao, Shandong, China}

%\affil[2]{\orgdiv{Department}, \orgname{Organization}, \orgaddress{\street{Street}, \city{City}, \postcode{10587}, \state{State}, \country{Country}}}

%\affil[3]{\orgdiv{Department}, \orgname{Organization}, \orgaddress{\street{Street}, \city{City}, \postcode{610101}, \state{State}, \country{Country}}}


\abstract{
This paper infers a single parameter in  non-sparse logistic regression models.  By transforming the null hypothesis into a moment condition, we construct the test statistic and obtain the asymptotic null distribution. Numerical experiments show that our method performs well.
}




\maketitle

\section{Introduction}\label{sec1}
The logistic regression models have been widely used in finance and genetics analysis,
which increasingly rely on high-dimensional observations. In other words, the dimension $p$  is high, and the sample size $n$ is relatively small, i.e. $n\rightarrow \infty$ and $p/n\rightarrow \infty$, therefore, modeling, inference, and prediction  become more challenging than in traditional environments.

%In this paper, we consider testing the high-dimensional logistic regression model:
%\begin{equation} \label{1.1}
%y_{i}=f(\beta^{T}X_{i})+\varepsilon_{i}, i=1,2,...n, \tag{1.1}
%\end{equation}
%where $f(u)=e^{u}/(1+e^{u})$ and $\varepsilon_{i}$ is error term, such that $E(\varepsilon_{i})=0$, $Var(\varepsilon_{i})=f(X_{i}^{T}\beta)(1-f(X_{i}^{T}\beta))$,
%we assume that  $\varepsilon$ and design matrix $X\in\mathcal{R}^{n\times p}$ are independent.
% The parameter vector $\beta\in\mathbb{R}^{p}$ is unknown and is  allowed  $p\gg n$.
%And we assume that  $y_{i}\mid X_{i} \thicksim Bernoulli(f(\beta^{T}X_{i}))$ independently for each $i=1,2,...,n$.
%In this paper, we focus on the fundamental problem of significance testing of single entries of the model parameter, i.e.
%\begin{equation}\label{1.2}
%H_{0}: \beta_{*}=\beta_{0}, \ \ versus \ \ H_{1}:\beta_{*}\neq\beta_{0}.   \tag{1.2}
%\end{equation}
%where  $\beta_{0}$ is a given value. Testing a single coefficient component is of great significance in practice and is a prerequisite for statistical inference.

Hypothesis test and confidence intervals in high-dimensional generalized linear models have been widely studied.
\cite{2014On} constructed confidence intervals and statistical tests for single or low-dimensional components of regression coefficients.
\cite{Ning0A2017} proposed a general framework for hypothesis testing and confidence intervals for low-dimensional components   based on general penalized M-estimators.
 \cite{2019Optimal} constructed a debiased estimator based on Lasso estimator and consistently established its asymptotic normality for future observations of arbitrary high dimensions.
 In the logistic regression models,   \cite{Sur2019} studied the likelihood ratio test under $p/n\rightarrow k$ for some $k<\frac{1}{2}$.
 \cite{Shi2021} focused on the logistic link and imposed certain stringent assumptions.
  \cite{2021RongMa} constructed a test statistic for testing the global null hypothesis using a generalized low-dimensional projection for bias correction.
   %proposed testing procedures for both the global null and the large-scale simultaneous hypotheses.
 % constructed a test statistic for testing the global null hypothesis using a generalized low-dimensional projection for bias correction and derived its asymptotic null distribution.
%Considering high-dimensional sparse logistic regression models for prediction,
\cite{Guo2021} proposed a novel bias-corrected estimator through  linearization and variance enhancement techniques.

The above methods are  sensitive to the sparsity assumption, which leads to the easy loss of error control when this assumption is violated. Statistical inference in non-sparse linear models has been studied extensively.
\cite{LuLin2011} proposed semiparametric re-modeling and inference method.
\cite{LuFeng2011} introduced a simulation-based procedure to reformulate a new model, with no need to estimate high-dimensional nuisance parameter.
\cite{Dezeure2017} proposed a residual and wild bootstrap methodology for individual and simultaneous inference.
By transforming the null hypothesis into a testable moment condition, \cite{2018Significance} proposed an asymptotically sparse CorrT method to solve the single-parameter testing problem.
By convolving the variables from the two samples and combining the moment method, \cite{2016Two} conducted the homogeneity test  of the global parameters in  two populations.
\cite{2016Linear}  further extended this moment method to test linear functionals of the regression parameters, and proposed Modified Dantzig Selector (MDS)  to estimate model parameters.
\cite{2022TESTABILITY} developed uniform and essentially uniform nontestability which  identified a collection of alternatives such that the power of any test was  at most equal to the nominal size.

In this paper, we consider  single parameter significance test in high-dimensional non-sparse logistic regression models, which is of great importance in practice, and is a prerequisite to statistical analysis.
%For example, the effects of treatment / drug on response are studied after controlling the effects of genetic markers with a high dimension.
For example, we study the effect of a treatment/drug on response after controlling for the impact of high-dimensional non-sparse genetic markers.
%In this paper, we consider the statistical inference of logistic regression models under the assumption of non-sparsity. In particular, we want to test  a single parameter.
This problem of statistical inference has not been solved in the existing literature.
  First, we  linearize  the regression function based on the logistic Lasso estimator.
   %This linearization technique helps to generalize the inference methods for linear models to logistic models.
Then, the approximate linear model is reconstructed according to the hypothesis, which is transformed into a testable moment condition.
Finally, we use MDS estimators  to construct the test statistics and prove the asymptotic null distribution and power property. Besides its applicability in logistic regression, this method can be extended to other nonlinear regression models.

%The main contribution of this paper is two-fold.\\
%(1) We propose a new method for single parameter testing in high dimensional logistic regression. The dimension $p$ is allowed to be much larger than the sample size $n$, and we do not make any assumptions about parameter sparsity. \\
%(2) For logistic regression, we first find the linearization of the regression function and then apply the method of moments.

%  Combining the methods in the above literature and the scenarios described,  we extend the reconstruction regression and convolutional regression methods to simultaneous inference of any group of parameters in the one- and two-sample cases.
% We transform the null hypothesis into moment conditions and construct  test statistics. The process does not rely heavily on accurate estimates of the parameters.
%  Without imposing any sparsity assumptions, the test statistics guarantee asymptotic control of type I error and Type II error.

  The remainder of this paper is organized as follows. In Section \ref{section 2}, We present a significance test method for single parameter in non-sparse logistic regression model, and introduce a new moment construction method.
   Section \ref{section 3}  shows the size and power properties of the proposed test.
    Section \ref{Numerical Examples} shows the numerical experiments and compares them with the results of another advanced method.
 %The complete details of the theoretical proofs are contained in supplement.

\section{Single parameter significance test} \label{section 2}
 \subsection{Notations}
 For a vector $V\in \mathbb{R}^{k}$,   $v_{i}$ represents the $i$-th element of $V$. $\|V\|_{\infty}=\max\limits_{1\leq i \leq k}\vert v_{i}\vert$ and $\|V\|_{0}=\sum\limits_{i=1}^{k}\mathrm{I}(v_{i}\neq 0)$, where $\mathrm{I}( \cdot )$ denotes the indicator function.
 For matrix $A$, its $(i,j)$ entry is denoted by $A_{i,j}$, and the $i$-th row is denoted by $A_{i}$.
 For two sequences $a_{n}, b_{n} >0$,  $a_{n}\asymp b_{n}$ means that there exist constants $C_{1}, C_{2} >0$ such that $\forall n$, $a_{n}\leq C_{1}b_{n}$ and $b_{n} \leq C_{2}a_{n}$.

 \subsection{Model and hypothesis}
%Consider the following high-dimensional logistic regression model
We consider the non-sparse logistic regression model:
\begin{equation} \label{1.1}
y_{i}=f(\beta^{T}X_{i})+\varepsilon_{i},i=1,2,...,n  \tag{2.1}
\end{equation}
%We assume the observations $(X_{i},y_{i})\in \mathbb{R}^{p}\times \{0,1\}, \ i=1,2,...,n$, are independently generated from
%%In this paper, we consider testing the  logistic regression model:
%\begin{equation} \label{1.1}
%y_{i}| X_{i}\sim Bernoulli(f(X_{i}^{T}\beta)),  \tag{2.1}
%\end{equation}
where $f(u)=e^{u}/(1+e^{u})$, and  $\beta=(\beta_{*}, \theta_{*})\in\mathbb{R}^{p}$ is a non-sparse regression vector with single parameter $\beta_{*}$ and redundant parameter $\theta_{*}\in\mathbb{R}^{p-1}$.
  The observations are i.i.d. samples $(X_{i},y_{i})\in \mathbb{R}^{p}\times \{0,1\}$ for $i=1,2,...,n$, and  $y_{i}\mid X_{i} \thicksim Bernoulli(f(\beta^{T}X_{i}))$ independently for each $i=1,2,...,n$. We assume  $X_{i}\sim N(0,\Sigma)$. In fact, this result can be extended to sub-Gaussian distribution.
%And $\varepsilon_{i}$ is error term, such that $E(\varepsilon_{i})=0$, $Var(\varepsilon_{i})=f(X_{i}^{T}\beta)(1-f(X_{i}^{T}\beta))$.
The $\varepsilon=(\varepsilon_{1}, \varepsilon_{2},..., \varepsilon_{n})^{'}\in\mathbb{R}^{n}$ is the error term, which is not correlated with $X=(X_{1},X_{2},...,X_{n})^{'}\in\mathbb{R}^{n\times p}$.

In this paper, we focus on the  significance test of single parameter $\beta_{*}$, i.e.
\begin{equation}\label{1.2}
H_{0}: \beta_{*}=\beta_{0}, \ \ versus \ \ H_{1}:\beta_{*}\neq\beta_{0}.   \tag{2.2}
\end{equation}
where  $\beta_{0}$ is a given value.
%our construction of the single entries of the model parameter testing procedure begins with a regularized estimator such as $\mathcal{L}_{1}$-regularized M-estimator. For high-dimensional logistic regression,
As a preliminary,  we first give an estimator of the global parameter $\beta$.
For technical reasons, we split the samples  into two independent subsets $\mathcal{D}_{1}$ and $\mathcal{D}_{2}$. The  $\mathcal{L}_{1}$-regularized M-estimator $\hat{\beta}$ of  $\beta$ is obtained from $\mathcal{D}_{1}$:
\begin{equation}\label{M-estimator}
\hat{\beta}=\arg\min\limits_{\beta}\{\frac{1}{n}\sum\limits_{i=1}^{n}[-y_{i}\beta^{T}X_{i}+log(1+e^{\beta^{T}X_{i}})]+\lambda\|\beta\|_{1}\},   \tag{2.3}
\end{equation}
which is the minimizer of a penalized log-likelihood function with $\lambda\asymp\sqrt{\frac{logp}{n}}$.
Although $\hat{\beta}$ can achieve the optimal rate of convergence \citep{Sahand2012A, 2012Estimation}, it's not suitable to construct confidence intervals and  hypotheses test directly because of its biases.

 In the following, we reconstruct the regression model based on $\hat{\beta}$ and the samples from $\mathcal{D}_{2}$.
% Throughout the paper, the samples $(X_{i},Y_{i}), i=1,2,...,n$, are from $\mathcal{D}_{2}$, which are independent of $\hat{\beta}$.
We consider the Taylor expansion of $f(u_{i})$ at $\hat{u}_{i}$ for $u_{i}=\beta^{T}X_{i}$ and $\hat{u}_{i}=\hat{\beta}^{T}X_{i}$,
\begin{equation}\label{Taylor expansion}
f(u_{i})=f(\hat{u}_{i})+\dot{f}(\hat{u}_{i})(u_{i}-\hat{u}_{i})+Re_{i}, \ i=1,2,...,n, \tag{2.4}
\end{equation}
where $\dot{f}(\cdot)$ is the derivative of $f(\cdot)$, and $Re(i)$ is the reminder term. Plugging \eqref{Taylor expansion} into  \eqref{1.1}, we have
\begin{equation}\label{moxingchongxie}
y_{i}-f(\hat{\beta}^{T}X_{i})+\dot{f}(\hat{\beta}^{T}X_{i})X_{i}^{T}\hat{\beta}-Re_{i}=\dot{f}(\hat{\beta}^{T}X_{i})X_{i}^{T}\beta+\varepsilon_{i}, \ i=1,2,...,n. \tag{2.5}
\end{equation}
We can treat $y_{i}-f(\hat{\beta}^{T}X_{i})+\dot{f}(\hat{\beta}^{T}X_{i})X_{i}^{T}\hat{\beta}-Re_{i}$ as a new response variable $y_{new,i}$, and $Y_{new}=(y_{new,1},y_{new,2},...,y_{new,n})^{T}\in \mathbb{R}^{n}$, whereas $\dot{f}(\hat{\beta}^{T}X_{i})X_{i}$ as the new covariate $X_{new,i}=(z_{i},W_{i}^{T})^{T}$ with $z_{i}\in\mathbb{R}$ and $W_{i}\in\mathbb{R}^{p-1}$. Consequently, $\beta$ can be considered as the regression coefficient of this approximate linear model. Then  \eqref{moxingchongxie} is transformed into
\begin{equation}\label{moxingchongxie1}
y_{new,i}=X_{new,i}^{T}\beta+\varepsilon_{i}, \ i=1,2,...,n. \tag{2.6}
\end{equation}

Since the null hypothesis is $H_{0}: \beta_{*}=\beta_{0}$, the above equation can be rewritten as:
\begin{equation}\label{moxingchongxie2}
y_{new,i}=z_{i}\beta_{*}+W_{i}^{T}\theta_{*}+\varepsilon_{i}, \ i=1,2,...,n,\tag{2.7}
\end{equation}
where $Z=(z_{1},z_{2},...,z_{n})^{'}\in\mathbb{R}^{n}$ and $W=(W_{1},W_{2},...,W_{n})^{'}\in\mathbb{R}^{n\times(p-1)}$  are the design matrices.
%And $\theta_{*}\in\mathfrak{R}^{p-1}$ is the vector of redundant parameters.
Subtracting $z_{i}\beta_{0}$ from both sides of model \eqref{moxingchongxie2}, we build the following reconstructed model
\begin{equation}\label{chonggoumoxing11111}
y_{new,i}-z_{i}\beta_{0}=z_{i}\gamma_{*}+W_{i}^{T}\theta_{*}+\varepsilon_{i}, \ i=1,2,...,n, \tag{2.8}
\end{equation}
where $\gamma_{*}=\beta_{*}-\beta_{0}$ is of main interest, and  the original $H_{0}$ in \eqref{1.2} is equivalent to
\begin{equation}\label{chonggoumoxing22222}
H_{0}:\gamma_{*}=0. \tag{2.9}
\end{equation}
Thus, we define a pseudo-response $V=Y_{new}-Z\beta_{0}$ and a pseudo-error $e=Z\gamma_{*}+\varepsilon$, which satisfy that
\begin{equation}\label{VWe}
V=W\theta_{*}+e, \tag{2.10}
\end{equation}
 Since  $X$ and $\varepsilon$ are unrelated, $W$  and $\varepsilon$ are unrelated. Therefore, when the null hypothesis is true, we have $E(e^{T}W)=E(\varepsilon^{T}W)=0$. Otherwise, $e$ and  $W$ may be linear dependent through $z$, which is caused by the confounding effects of $W$ and $Z$.
Next, we establish a linear correlation model between $Z$ and $W$:
\begin{equation}\label{xianxingxiangguanmoxing}
Z=W\pi+U, \tag{2.11}
\end{equation}
where $\pi\in\mathbb{R}^{p-1}$ is an unknown regression coefficient vector and $U\in\mathbb{R}^{n}$ is the error term,  internally independent of each other, which  follows Gaussian distribution with zero mean, and $U$ is uncorrelated to $(V, \ W)$.
It is worth mentioning that we assume that $\pi$ is sparse to decouple the correlation between $Z$ and $W$.
%We assume that $\pi$ is sparse,  to decouple the dependence between $Z$ and $W$.
%Sparse $\pi$ is a generalization of the sparsity condition on the precision matrix $\Omega_{X}=\Sigma_{X}^{-1}$, which is a typical regularity condition.

We consider the correlation between  $e$  in \eqref{VWe} and $U$ in \eqref{xianxingxiangguanmoxing}:
\begin{equation}\label{liangwuchaxiangguanxing}
E(U^{T}e)=E(U^{T}Z\gamma_{*}+U^{T}\varepsilon)=E(U^{T}U)\gamma_{*}. \tag{2.12}
\end{equation}
Therefore, the original test problem in \eqref{1.2} is equivalent to
\begin{equation}\label{jianyanwenti}
H_{0}: E((V-W\theta_{*})^{T}(Z-W\pi))=0 \ \ versus \ \ H_{1}: E((V-W\theta_{*})^{T}(Z-W\pi))\neq0. \tag{2.13}
\end{equation}
Since $\pi$ is sparse,  consistent estimator $\breve{\pi}$ is easy to obtain. However, it is difficult to obtain the consistent estimator of non-sparse parameter $\theta_{*}$.
For any estimator $\check{\theta}_{*}$ of $\theta_{*}$,  we have
\begin{equation}
E((V-W\breve{\theta}_{*})^{T}(Z-W\breve{\pi}))\rightarrow E((V-W\theta_{*})^{T}(Z-W\pi))+E((Z-W\pi)^{T}W(\theta_{*}-\breve{\theta}_{*})). \notag
\end{equation}
In the above equation, $\breve{\theta}_{*}$ is a function of $(V,W)$, while $U$ is uncorrelated to $(V,W)$,  so $Z-W\pi$ and $W(\theta_{*}-\breve{\theta}_{*})$ are uncorrelated. Then
\begin{equation}
E((Z-W\pi)^{T}W(\theta_{*}-\breve{\theta}_{*}))=E(Z-W\pi)^{T}E(W(\theta_{*}-\breve{\theta}_{*}))=0. \notag
\end{equation}
Therefore,
\begin{equation}
E((V-W\breve{\theta}_{*})^{T}(Z-W\breve{\pi}))\rightarrow E((V-W\theta_{*})^{T}(Z-W\pi)). \notag
\end{equation}
The  inner product structure in  \eqref{jianyanwenti} alleviates the reliance on a good estimator of $\theta_{*}$.
We will estimate the unknown parameters $\pi$ and $\theta_{*}$ in the next subsection.
 %The moment condition in  \eqref{liangwuchaxiangguanxing} contains the sparse parameter $\pi$ and the dense parameter $\theta$.

\subsection{Modified Dantzig Selector} \label{section 2.2}
 MDS is used  to estimate   the unknown parameter $\theta_{*}$  and error variance $\sigma_{e}^{2}$   simultaneously,

%In the existing literature, there are many methods that can be used to fit sparse regression models, such as Lasso \citep{Robert1996Regression}, Dantzig selector \citep{Cand2007Rejoinder} and so on. However, in the above estimation methods, the choice of tuning parameters $\lambda$ deeply affects the performance of the estimators.
%Some empirical and theoretical studies suggest that the tuning parameters should be proportional to the  noise standard deviation $\sigma_{\varepsilon}$, i.e. $\lambda=\sigma_{\varepsilon}\sqrt{log p/n}$. Unfortunately, noise variance is unavailable in most applications.
% Therefore, estimating unknown parameters together with noise variance in a joint fashion is important. This estimation methods include square-root Lasso \citep{2011Square}, Scaled Lasso \citep{2012Scaled}, Self-tuned Dantzig Selector \citep{2013Pivotal} and so on.
%However, they didn't provide reasonable estimators under non-sparse and high-dimensional models in which it is impossible to construct consistent estimators of model parameters. Constrained by the disadvantages of the above methods, we utilize the Modified Dantzig selector (MDS) \citep{2016Linear} under non-sparse conditions which has the advantage of being well controlled in certain sense.
% The MDS estimator of $\theta_{*}$ is defined as follows
 \begin{equation}\tag{2.14} \label{theta_tilde}
\begin{split}
\tilde{\theta}_{*}=&\arg\min\limits_{\theta_{*}\in\mathbb{R}^{p-1}} \|\theta_{*}\|_{1} \notag
\\
s.t.& \ \|W^{T}(V-W\theta_{*})\|_{\infty}\leq \eta\rho_{1}\sqrt{n}\|V\|_{2}
\notag \\
&V^{T}(V-W\theta_{*})\geq \rho_{0}\rho_{1}\|V\|_{2}^{2}/2 \notag \\
&\rho_{1} \in [\rho_{0}, 1] \notag,
\end{split}
\end{equation}
where  $\rho_{1}=\sigma_{e}/ \sqrt{E(v_{1})^{2}}$, and $\rho_{0} \in (0,1)$ is a lower bound for this ratio.
$\eta\asymp \sqrt{n^{-1}log p}$ is a tuning parameter.

Similarly, the estimator $\tilde{\pi}\in \mathbb{R}^{p-1}$ of   $\pi$ is
\begin{equation} \tag{2.15} \label{pij_tilde}
\begin{split}
\tilde{\pi}=&\arg\min\limits_{\pi\in\mathbb{R}^{p-1}} \|\pi\|_{1}
\notag \\
s.t.& \|W^{T}(Z-W\pi)\|_{\infty}\leq\eta\rho_{2}\sqrt{n}\|Z\|_{2}
\notag \\
&Z^{T}(Z-W\pi)\geq \rho_{0}\rho_{2}\|Z\|_{2}^{2}/2 \notag \\
& \rho_{2}\in[\rho_{0},1] \notag,
\end{split}
\end{equation}
where $\rho_{2}=\sigma_{u}/\sqrt{E(z_{1})^{2}}$.

\subsection{Test statistic} \label{ section 2.3}
% In this subsection, we construct a test statistic  using $\tilde{\theta}_{*}$ and $\tilde{\pi}$  to test the moment condition in \eqref{liangwuchaxiangguanxing}.

 By plugging in the estimators $\tilde{\pi}$ and $\tilde{\theta}_{*}$, we construct the following test statistic
 \begin{equation}\label{jianyantongjiliang}
T_{n}= n^{-\frac{1}{2}}\hat{\sigma}_{e}^{-1}(Z-W\tilde{\pi})^{T}(V-W\tilde{\theta}_{*})
\tag{2.16},
\end{equation}
where $\hat{e}=V-W\tilde{\theta}_{*}$ and  $\hat{\sigma}_{e}=\|V-W\tilde{\theta}_{*}\|_{2}/\sqrt{n}$.
Obviously,  under the null hypothesis and the sparsity assumption of $\pi$,  we have
\begin{equation}
T_{n}=n^{-\frac{1}{2}}\hat{\sigma}_{e}^{-1}(Z-W\tilde{\pi})^{T}(V-W\tilde{\theta}_{*}) =
\Delta\cdot\hat{\sigma}_{e}^{-1}+n^{-\frac{1}{2}}\hat{\sigma}_{e}^{-1}U^{T}\hat{e} \notag,
\end{equation}
where $\Delta=n^{-\frac{1}{2}}(\pi-\tilde{\pi})^{T}W^{T}\hat{e}$, and we can proof that
 $\Delta\cdot\hat{\sigma}_{e}^{-1}=o_{p}(1)$.
So the statistical properties of  $T_{n}$ is determined by $n^{-\frac{1}{2}}U^{T}\hat{e}\hat{\sigma}_{e}^{-1}=n^{-\frac{1}{2}}\sum\limits_{i=1}^{n}u_{i}\hat{e}_{i}\hat{\sigma}_{e}^{-1}$.

%Our approximation method depends on the implicit independence of $U$ and $\hat{e}$.
Under the null hypothesis,  $U$ is uncorrelated of $(V, W)$, while   $\tilde{\theta}_{*}$  is completely dependent on $(V,W)$, so $\hat{e}$ is also only related to $(V,W)$. Therefore, $U$ and $\hat{e}$ are independent. Because of this independence, we have
\begin{align}
E[n^{-\frac{1}{2}}\sum\limits_{i=1}^{n}u_{i}\hat{e}_{i}\hat{\sigma}_{e}^{-1}\mid\hat{e}_{i}]=&\frac{1}{\|\hat{e}\|_{2}}\sum\limits_{i=1}^{n}\hat{e}_{i}E(u_{i})=0 \notag, \\
Var[n^{-\frac{1}{2}}\sum\limits_{i=1}^{n}u_{i}\hat{e}_{i}\hat{\sigma}_{e}^{-1}\mid\hat{e}_{i}]
=&n^{-1}\hat{\sigma}_{e}^{-2}\sum\limits_{i=1}^{n}\hat{e}_{i}^{2}Var(u_{i})
=E(u_{1}^{2}) \notag.
\end{align}
%where $u_{i}$ is the $i$-th element of $U$.
Therefore, according to the Gaussianity of $U$,
the distribution of $n^{-\frac{1}{2}}\sum\limits_{i=1}^{n}u_{i}\hat{e}_{i}\hat{\sigma}_{e}^{-1}$ conditional on  $\{\hat{e}_{i}\},i=1,2,...,n$ is $N(0,Q)$ and $Q=E(u_{1}^{2})$. That is,
\begin{equation}
n^{-\frac{1}{2}}\hat{\sigma}_{e}^{-1}\hat{e}^{T}U\mid \hat{e}\sim N(0, Q ),
\notag
\end{equation}
where $Q$ is unknown, which we want to replace with a natural estimator $\hat{Q}=\frac{1}{n}\sum\limits_{i=1}^{n}\hat{u}_{i}^{2}$.
%And we can proof that $|\hat{Q}-Q|=\sqrt{n^{-1}logp}$.
%We consider  a natural estimator $\hat{Q}=\frac{1}{n}\sum\limits_{i=1}^{n}\hat{u}_{i}^{2}$,
%where $\hat{u}_{i}$  is the $i$-th element of $\hat{U}=Z-W\tilde{\pi} \in \mathbb{R}^{n}$.
%We introduce the function $\Gamma(x;A)=P(\mid \xi\mid\leq x)$ with $\xi\sim N(0,A)$.
%And we will show that  $\Gamma(x;\hat{Q})=\mid N(0,\hat{Q})\mid$ can  asymptotically approximate the distribution of  $T_{n}$. The critical value of $\Gamma(x;\hat{Q})$ can be easily computed by simulation.

\section{Theoretical results} \label{section 3}
\subsection{Size property}
 %We now turn to the properties of the introduced test while imposing extremely weak conditions.
 We now turn our attention to the property of the test, which is imposed under extremely weak conditions.
\begin{assumption} \label{assumption 2.1}
Consider the model \eqref{1.1}. Suppose that the following hold: \\
%(i) the design matrix $X$ follows a Gaussian distribution, that is $X\sim N(0,\Sigma)$;\\
(i)  there exist constants $c$, $d$ $\in (0,+\infty)$ such that the eigenvalues of covariance matrix $\Sigma$ lie in $(c,d)$;\\
(ii) $\pi$ is sparse, which means $s_{\pi}=o(\sqrt{n/log^{3}p})$, where
$s_{\pi}={\|\pi\|_{0}}$. \\
%(iv) $s_{\pi}\|\beta_{G}-\beta_{G}^{0}\|_{0}+\|\theta\|_{0}=o(\sqrt{n}/logp)$; \\
%(iv') $\|\beta_{G}-\beta_{G}^{0}\|_{0}=o_{p}(\sqrt{logp})$ and $\|\theta\|_{0}=o(\sqrt{n}/logp)$. \\
%(v) there exist constants $\delta \  and\  \kappa_{1} \in (0,+\infty)$ such that: $E\vert \varepsilon \vert^{2+\delta}<\kappa_{1}$;\\
%(vi) there exist constants $K_{1} \ and \ K_{2} >0$, which only depends on the constants $\delta \  and\  \kappa_{1} $, satisfying that
% $ \|\Sigma_{u}(\beta_{G}-\beta_{G}^{0})\|_{\infty}\geq\sqrt{n^{-1}logp}(K_{1}\|\beta_{G}-\beta_{G}^{0}\|_{2}+K_{2}) $.
\end{assumption}
Assumption  \ref{assumption 2.1} is reasonably weak.
 Assumption  \ref{assumption 2.1}(i)  is  a common condition imposed in high-dimensional literature.
 Assumption \ref{assumption 2.1}(ii) imposes a sparsity condition on the regression coefficient vector $\pi$, rather than on $\beta$ or $\Sigma$ of the  model \eqref{1.1}, which shows that the following conclusions are robust to dense models.
Then  we provide the following result for  $T_{n}$.
\begin{theorem} \label{theorem 2.1}
 Let Assumption \ref{assumption 2.1} be  hold, when $n, p\rightarrow \infty$ with $logp=o(\sqrt{n})$, then under null hypothesis,
 \begin{align} \label{test}
  P(\vert T_{n}\vert>\hat{Q}^{\frac{1}{2}}\Phi^{-1}(1-\alpha/2))\rightarrow\alpha, \forall\alpha\in(0,1)
 \tag{3.1},
 \end{align}
 where $\Phi^{-1}(1-\alpha/2)$  is the $1-\alpha/2$ quantile of standard normal distribution.
\end{theorem}
Theorem \ref{theorem 2.1} shows that $T_{n}$, under the null hypothesis, converges  to $N(0,\hat{Q})$. Hence, a test with nominal size $\alpha\in(0,1)$ rejects null hypothesis if and only if $\vert T_{n}\vert>\hat{Q}^{\frac{1}{2}}\Phi^{-1}(1-\alpha/2)$.
%Theorem \ref{theorem 2.1} formally establishes that the new test is asymptotically exact in testing $\beta_{*}=\beta_{0}$.
In particular, the test is robust to dense $\theta_{*}$, in the sense that even under dense $\theta_{*}$, our procedure does not generate false positive results.
%In contrast to existing methods, we do not make any sparsity assumptions about the model parameter $\beta$.
 Instead of an inference on the basis of an estimator, it is a direct statistical conclusion on the basis of a null hypothesis.
 At the same time, we can construct confidence sets for $\beta_{*}$ even when the nuisance parameter $\theta_{*}$ is non-sparse.

%We can construct confidence sets for $\beta_{*}$ by inverting the above test. Let $1-\alpha$ be the nominal coverage level. We define
% \begin{align} \label{confidence interval}
% \mathcal{C}_{1-\alpha/2}:=\{\beta: \mid T_{n}\mid \leq \hat{Q}^{-\frac{1}{2}}\Phi^{-1}(1-\alpha/2)\} \tag{3.2},
% \end{align}
% Theorem \ref{theorem 2.1} guarantees the validity of the confidence set.

\begin{corollary} \label{corollary}
Let Assumption \ref{assumption 2.1} be hold and $1-\alpha$ be the nominal coverage level. We define
 \begin{align} \label{confidence interval}
 \mathcal{C}_{1-\alpha/2}:=\{\beta: \vert T_{n}\vert \leq \hat{Q}^{\frac{1}{2}}\Phi^{-1}(1-\alpha/2)\} \tag{3.2},
 \end{align}
 which has the exact coverage asymptotically:
 \begin{align} \label{coverage}
\lim\limits_{n,p\rightarrow\infty}P(\beta_{*}\in\mathcal{C}_{1-\alpha})=1-\alpha  \tag{3.3}.
 \end{align}
\end{corollary}

\subsection{Power property}
To evaluate the power property of the test,  we consider the following test problem:
 \begin{align} \label{H1}
H_{0}:\beta_{*}=\beta_{0} \ \ versus \ \ H_{1}:\beta_{*}=\beta_{0}+h  \tag{3.4},
 \end{align}
 where $h$ is a given constant.
 %It is clear that the difficulty of differentiating $H_{0}$ from $H_{1}$ depends on the magnitude of $h$.
 It is clear that the difficulty in distinguishing $H_{0}$ from $H_{1}$ depends on $h$.
 \begin{assumption} \label{assumption 2.2}
 Let Assumption \ref{assumption 2.1} be hold. In addition, suppose
 \\
%(i) the design matrix $X$ follows a Gaussian distribution, that is $X\sim N(0,\Sigma)$;\\
%(ii)  there exist constants $c$, $d$ $\in (0,+\infty)$ such that the eigenvalues of covariance matrix $\Sigma$ lie in $(c,d)$;\\
%(iii)  the $\pi$ is sparse, which means $s_{\pi}=o(\sqrt{n/log^{3}p})$, where
%$s_{\pi}={\\vert\pi\\vert_{0}}$; \\
(i) $\|\theta_{*}\|_{0}=o(\sqrt{n}/logp)$; \\
%(ii) $\|\beta_{G}-\beta_{G}^{0}\|_{0}=o_{p}(\sqrt{logp})$ and $\|\theta\|_{0}=o(\sqrt{n}/logp)$. \\
(ii) there exist constants $\delta \  and\  \kappa_{1} \in (0,+\infty)$ such that $E\vert \varepsilon \vert^{2+\delta}<\kappa_{1}$.
%(iv) there exist constants $K_{1} \ and \ K_{2} >0$, which only depends on the constants $\delta \  and\  \kappa_{1} $, satisfying that
 %$ \|\Sigma_{u}(\beta_{G}-\beta_{G}^{0})\|_{\infty}\geq\sqrt{n^{-1}logp}(K_{1}\|\beta_{G}-\beta_{G}^{0}\|_{2}+K_{2}) $.
\end{assumption}
Assumption \ref{assumption 2.2} is relatively mild.  The sparsity condition of $\theta_{*}$ is used to guarantee the asymptotic power of high-dimensional tests in Assumption \ref{assumption 2.2}(i), which implies the sparsity of the model, and it is consistent with the traditional test \citep{2013Two,2014On}.
 Assumption \ref{assumption 2.1}(ii) is a regular moment condition.
% For the convenience of  proof, we impose a $l_{\infty}$ norm lower bound on the product of $\Sigma_{u}$ and the deviation in Assumption \ref{assumption 2.1}(vi).
Then  we provide the following result for  $T_{n}$.

\begin{theorem} \label{theorem 2.2}
  Let $H_{1}$ in \eqref{H1} and Assumption \ref{assumption 2.2} be hold. When $n, p\rightarrow \infty$, with $logp=o(\sqrt{n})$, then  there exist constants $K_{1},K_{2}>0$ depending only on the constants in Assumption \ref{assumption 2.2} such that, whenever
  \begin{align}
\vert \sigma_{u}^{2}(\beta_{*}-\beta_{0})\vert\geq \sqrt{n^{-1}logp}(K_{1}\vert\beta_{*}-\beta_{0}\vert+K_{2}), \notag
 \end{align}
  where $\sigma_{u}^{2}=E(u^{2})$,  the test  is asymptotically  powerful, $i.e.$
 \begin{align}
 P(\vert T_{n}\vert>\hat{Q}^{\frac{1}{2}}\Phi^{-1}(1-\alpha/2))\rightarrow 1, \notag \ {\forall} \alpha \in (0,1).
 \end{align}
 \end{theorem}
 Theorem \ref{theorem 2.2} establishes the power property of the proposed test under the sparse model.

\section{Numerical Examples} \label{Numerical Examples}
In this section, we evaluate the proposed method in the finite sample setting by observing its behavior  in both simulated and real data.
\subsection{Simulation Examples}\label{section 4.1}
%In this subsection, we evaluate the proposed method in the finite-sample setting by observing its behavior on the simulated data setting.
We consider model \eqref{1.1}.
In all simulations, we set $n=200$, $p=500$ and the nominal size  is 5\%. The rejection probabilities are based on 100 repetitions. For application purposes, we recommend choosing the tuning parameters as $\eta=0.5\sqrt{\frac{log p}{n}}$ and $\rho_{0}=0.01$,  which are  commonly used options, and we will demonstrate in our simulations that it provides good results.
%\subsection{One-Sample Experiments }

For the test problem \eqref{1.2}, without loss of generality, we consider the test of the first component of the parameter, i.e.
\begin{equation} \label{simulatetest}
 H_{0}: \beta_{1}=\beta_{1}^{0} \ versus \ H_{1}: \beta_{1}=\beta_{1}^{0}+h  \tag{4.1},
\end{equation}
 where $\beta_{1}^{0}$ is a given constant.
  We show the results for three different Gaussian designs as follows. \\
(1) (Toeplitz) Here we consider the standard Toeplitz  design where the rows of $X$ are drawn as an i.i.d random draws from a multivariate Gaussian distribution $N(0,\Sigma_{X})$, with covariance matrix $(\Sigma_{X})_{i,j}=0.4^{\vert i-j \vert}$. \\
(2) (Noncorrelation) Here we consider uncorrelated design where the rows of $X$ are i.i.d draws from  $N(0,\Sigma_{X})$, where $(\Sigma_{X})_{i,j}$ is 1 for $i=j$ and is 0 for $i\neq j$. \\
(3) (Equal correlation) Here we consider a non-sparse design matrix with equal correlation among the features. Namely, the rows of $X$ are i.i.d draws from  $N(0,\Sigma_{X})$, where $(\Sigma_{X})_{i,j}$ is 1 for $i=j$ and is 0.01 for $i\neq j$.

 %We also consider two specifications of the error distribution. \\
%(a) In the light-tailed case, $\varepsilon$ is taken from a standard normal distribution.\\
%(b) In the heavy-tailed case, the $\varepsilon$ is taken from Student's t-distribution with 3 degrees of freedom.

Let $s=\|\beta\|_{0}$ denotes model sparsity.
%The confidence intervals  proposed by \cite{2021TTonyCai}  are shown to achieve the optimal expected length up to a logarithmic factor over the sparse regime with $s=o(\frac{n}{logn logp})$. And Theorem \ref{theorem 2.1} shows that even under dense parameter $\beta$, our test proposed in \eqref{test}  is still valid.
 To show the size property of our method for dense model, we vary $s$ from  $s=10$ to extremely large $s=p$. For sparsity $s$, we set the model parameters as
$\beta_{j}=\frac{3}{\sqrt{p}}$, $1\leq j \leq s$ and $\beta_{j}=0$, $j>s$.
%We test the hypothesis $H_{0}: \beta_{1}=\frac{3}{\sqrt{p}}$ and
%the rejection probability represents the Type I error.

We compare our method with the generalized low-dimensional projection (LDP) method for bias correction \citep{2021RongMa}.
 The size results are collected in Table \ref{Table 1},
% where "Toeplitz" indicates Toeplitz design described in (1), "Noncorrelation" indicates uncorrelated design described in (2) and "Equal correlation" indicates equal correlated design described in (3).
%where LDP refers to the method proposed by \cite{2021RongMa} and Ours refers to the method proposed in this paper.
 where we can clearly see that the LDP method does not have the size property in the dense model, that is, the Type I error probabilities are much higher than the nominal level $\alpha$. This indicates that the LDP method  fails to dense models.
Conversely, when the sparsity of the model is equal to $s=p$, the Type I error probability of our method remains stable. That is true  even if we change the correlation among the features.
\begin{table}[h] \centering
\begin{center}
\begin{minipage}{\textwidth}
\caption{Size properties of LDP and our method}\label{Table 1}
\begin{tabular*}{\textwidth}{@{\extracolsep{\fill}}lcc|cc|cc@{\extracolsep{\fill}}}
\toprule%
& \multicolumn{2}{@{}c@{}}{Toeplitz} & \multicolumn{2}{@{}c@{}}{Noncorrelation} & \multicolumn{2}{@{}c@{}}{Equal correlation} \\ \cmidrule{2-3}\cmidrule{4-5}\cmidrule{6-7}%
Method & LDP       & Ours             & LDP           & Ours                & LDP            & Ours           \\
\midrule
s=10  & 0.70         & 0.09         & 0.66           & 0.05               & 0.61            & 0.05                \\
s=20  & 0.69         & 0.02         & 0.69           & 0.03               & 0.65            & 0.03                \\
s=50  & 0.72         & 0.02         & 0.70           & 0.02               & 0.79            & 0.05                \\
s=100 & 0.81         & 0.03         & 0.72           & 0.03               & 0.77            & 0.05                \\
s=n   & 0.82         & 0.05         & 0.78           & 0.07               & 0.89            & 0.05                \\
s=p   & 0.90         & 0.04         & 0.89           & 0.04               & 0.86            & 0.07                \\
\bottomrule
\end{tabular*}
\end{minipage}
\end{center}
\end{table}

For the first parameter component $\beta_{1}=\frac{3}{\sqrt{p}}$, we construct its $1-\alpha$ confidence intervals for different sparsity levels, and obtain the coverage probabilities (CP) based on 100 repetitions.
According to Theorem \ref{theorem 2.1}, the asymptotic distribution of $T_{n}$ is $N(0,\hat{Q})$. Also by the analysis in Section \ref{ section 2.3}, we have
\begin{align}
T_{n}\rightarrow n^{-\frac{1}{2}}\hat{\sigma}_{e}^{-1}U^{T}\hat{e}\rightarrow N(0,\hat{Q}), \notag
\end{align}
By inverting the solution $|T_{n}| \leq \hat{Q}^{\frac{1}{2}}\Phi^{-1}(1-\alpha/2)$, the $1-\alpha$ confidence interval of the parameter $\beta_{1}$ can be obtained as
%\beta_{1}^{0}\pm \frac{n^{\frac{1}{2}}\hat{Q}^{\frac{1}{2}}\Phi^{-1}(1-\alpha/2)\hat{\sigma}_{e}+\mid \hat{U}^{T}(W(\theta_{*}-\tilde{\theta}_{*})+\varepsilon)\mid}{\mid \hat{U}^{T}Z\mid}
\begin{align}
[\beta_{1}^{0}-\frac{n^{\frac{1}{2}}\hat{Q}^{\frac{1}{2}}\Phi^{-1}(1-\alpha/2)\hat{\sigma}_{e}+\hat{U}^{T}(W(\theta_{*}-\tilde{\theta}_{*})+\varepsilon)}{\hat{U}^{T}Z}, \notag\\
\beta_{1}^{0}+\frac{n^{\frac{1}{2}}\hat{Q}^{\frac{1}{2}}\Phi^{-1}(1-\alpha/2)\hat{\sigma}_{e}-\hat{U}^{T}(W(\theta_{*}-\tilde{\theta}_{*})+\varepsilon)}{\hat{U}^{T}Z}]. \notag
\end{align}
%In addition, in \cite{2021RongMa}, since
%$
%M_{1}=\frac{\check{\beta}_{1}-\beta_{1}}{\tau_{1}}\rightarrow N(0,\sigma^{2})
%$, the confidence interval of parameter $\beta_{1}$ is
%\begin{align}
%\check{\beta}_{1}\pm \sigma\tau_{1}\Phi^{-1}(1-\alpha/2). \notag
%\end{align}
The results for confidence intervals (CI), lengths and CP are collected in Table \ref{Table 222}.
\begin{table}[h]
\begin{center}
\begin{minipage}{\textwidth}
\caption{Confidence intervals, lengths and coverage probabilities}\label{Table 222}
\begin{tabular*}{\textwidth}{@{\extracolsep{\fill}}lccc|ccc|ccc@{\extracolsep{\fill}}}
\toprule%
  & \multicolumn{3}{c}{Toeplitz}      & \multicolumn{3}{c}{Noncorrelation}                 & \multicolumn{3}{c}{Equal correlation} \\ \cmidrule{2-4}\cmidrule{5-7}\cmidrule{8-10}%
Sparsity & CI             & Length & \multicolumn{1}{c|}{CP} & CI              & Length & \multicolumn{1}{c|}{CP} & CI                & Length   & CP     \\ \hline
s=10     & (-0.2,0.5) & 0.7  & 94\%                    & (-0.3,0.3) & 0.6  & 95\%                    & (-0.3,0.4)    & 0.7    & 95\%   \\
s=20     & (-0.2,0.4) & 0.6  & 94\%                    & (-0.2,0.4)  & 0.6  & 96\%                    & (-0.2,0.5)    & 0.7    & 95\%   \\
s=50     & (-0.2,0.5) & 0.7  & 93\%                    & (-0.3,0.3)  & 0.6  & 98\%                    & (0.3,0.3)     & 0.6    & 95\%   \\
s=100    & (-0.4,0.2) & 0.6  & 97\%                    & (0,0.5)  & 0.5  & 94\%                    & (-0.4,0.2)    & 0.6    & 95\%   \\
s=n      & (-0.1,0.6) & 0.7  & 99\%                    & (-0.2,0.3)  & 0.5  & 95\%                    & (-0.3,0.3)    & 0.6   & 91\%   \\
s=p      & (-0.4,0.2) & 0.6  & 94\%                    & (-0.2,0.4)  & 0.6  & 95\%                    & (0,0.6)     & 0.6   & 95\%   \\
\bottomrule
\end{tabular*}
\end{minipage}
\end{center}
\end{table}

\begin{figure}[h]\centering
\includegraphics[width=0.8\textwidth]{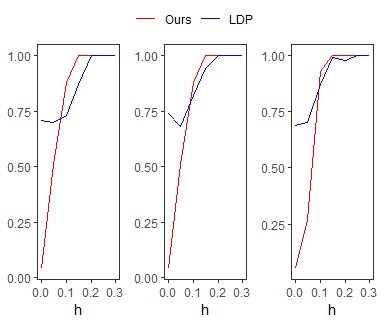}
\caption{Power curves of competing methods under different settings of design matrix}\label{111}
\end{figure}

In addition, Theorem \ref{theorem 2.2} gives the power property of the test under  sparse models ($\|\theta_{*}\|_{0}=o(\sqrt{n}/logp)$). For simplicity, we observe the power property only for $s=3$.
 The data is generated by the same model as in Table \ref{Table 1}, except that the true value of $\beta_{1}=\frac{3}{\sqrt{p}}+h$.
The results are collected in Figure \ref{111}, which presents full power curves with various values of $h$. Therefore, the far left presents  Type I error ($h=0$)  whereas other points on the curves correspond to Type II error ($h\neq 0$).
%In all the power curves, we use a sparse model with sparsity $3$, $i.e.$   $\|\beta\|_{0}=3$.
%In figure \ref{111},  the far left is the Toeplitz design, the middle is the unrelated design, and the far right is the equal-related design.
 We clearly observe that our method outperforms LDP by providing firm Type I error and reaching full power quickly.
%
%
%
%
%the first row corresponds to the light-tailed error distribution, and the second row corresponds to the heavy-tailed error distribution.
%The two plots in the first column correspond to the Toeplitz designs, where we clearly observe that the MMDS outperforms both MDL and WDL by providing firm Type I error probability and also reaching full power quickly.
%The second column corresponds to the uncorrelated design. The MDL and WDL  have  type I error probabilities over 0.05 whereas the MMDS still provides valid inference.
%The last column corresponds to the design with equal correlation. The MDL and WDL  completely break down with Type I error probability being close to 1.
%In conclusion, our method are stable in different settings of design matrices and model errors,
%However, the LDP method cannot control the probabilities of Type I error or Type II error.
Therefore, our proposed method provides a robust and more broadly applicable alternative to the existing inference process, achieving better error control.
\subsection{Real Data}\label{section 4.2}
We illustrate our proposed method by analyzing  "Lee Silverman voice treatment" (LSVT) voice rehabilitation dataset \citep{Athanasios2014Objective}. Vocal performance degradation is a common symptom for the vast majority of Parkinson's disease (PD) subjects. The current study aims to investigate the potential of automatically assessing sustained vowel articulation as  “acceptable” (a clinician would allow persisting in speech treatment) or “unacceptable” (a clinician would not allow persisting in speech treatment). We first standardized the data. The complete data includes 309 dysphonia measures, where each produces a single number per phonation, resulting in a design matrix of size $126\times 309$. There are no missing entries in the design matrix. This is a high-dimensional logistic regression problem with $n=126$ and $p=309$. We try to determine "which of the originally computed dysphonia measures matter in this problem."

Results are reported in Table \ref{Table3}. Therein we report the significant variables identified using our approach and LDP that affect the assessments of speech experts, respectively.
%A total of 11 significant variables were selected by our method.
%where $x_{3}$, $x_{18}$, $x_{37}$, $x_{97}$, $x_{100}$, $x_{111}$, $x_{115}$, $x_{229}$, $x_{230}$, $x_{231}$, $x_{265}$ represent dysphonia measures $F0-PQ5-classical_Schoentgen$, $pitc-PQ5-classical-Schoentgen$, $Ampl-PQ5-classical-Baken$, $delta log energy$, $2nd delta$, $delta delta log energy$, $3rd delta-delta$, $Ed2-9-coef$, $Ed2-10-coef$, $entropy-shannon3-1-coef$, $det-TKEO-std3-5-coef$,  respectively.
In addition to the above 11 dysphonia measures, the LDP method  selects 98 measures  as significant variables.
\begin{center}
\begin{table}[]\centering
\caption{Significant variables selected by our method and the LDP method} \label{Table3}
\begin{tabular}{ccc}
\hline
     & Dysphonia Measure & Number \\ \hline
Ours & $x_{3}$, $x_{18}$, $x_{37}$, $x_{97}$, $x_{100}$, $x_{111}$, $x_{115}$, $x_{229}$, $x_{230}$, $x_{231}$, $x_{265}$                  & 11     \\
LDP  & the above + $x_{4}$, $x_{4}$, $x_{6}$, $x_{7}$, $x_{8}$, $x_{9}$...                 & 109    \\ \hline
\end{tabular}
\end{table}
\end{center}

We divide the 126 samples into two parts, in which the first 100 samples are used as the training set and the last 26 samples are used as the testing set.  The significant variables selected by the two methods are used to fit the logistic regression model on the training set.
The logistic regression model obtained by our method is

\begin{align}
\hat{f}_{Ours}=\frac{e^{-1.8+52.8x_{3}+...+0.45x_{231}-25.73x_{265}}}{1+e^{-1.8+52.8x_{3}+...+0.45x_{231}-25.73x_{265}}}. \notag
\end{align}
And  LDP's logistic regression model  is
\begin{align}
\hat{f}_{LDP}=\frac{e^{-171.4-918.1x_{3}+...-14563x_{264}+30204x_{265}}}{1+e^{-171.4-918.1x_{3}+...-14563x_{264}+30204x_{265}}}. \notag
\end{align}
 The predicted values $y_{i}\mid X_{i}\sim Bernoulli(\hat{f}), i=1,2,...,26$, which are shown in Table \ref{Table4}.
%\begin{center}
%\begin{table}[] \label{Table4}
%\caption{The predicted values of our method and the LDP method} \label{Table4}
%\begin{tabular}{ccccccccccc}
%\hline
%Measure         & 1    & 2    & 3    & 4    & 5    & 6    & 7    & 8    & 9    & 10   & 11   & 12   & 13   \\ \hline
%Original Value  & 0    & 0    & 1    & 0    & 0    & 1    & 0    & 0    & 1    & 0    & 0    & 1    & 0    \\
%Ours            & 1    & 0    & 0    & 0    & 1    & 1    & 1    & 1    & 1    & 0    & 0    & 1    & 1 \\
%LDP             & 0    & 1    & 0    & 0    & 0    & 1    & 1    & 0    & 1    & 0    & 1    & 0    & 0  \\ \hline
%Measure         & 14   & 15   & 16   & 17   & 18   & 19   & 20   & 21   & 22   & 23   & 24   & 25   & 26 \\ \hline
%Original  Value & 0    & 1    & 0    & 0    & 1    & 0    & 0    & 1    & 0    & 0    & 1    & 0    & 0\\
%Ours            & 0    & 1    & 0    & 1    & 1    &0     &0     & 1 & 0 & 0    & 0 & 0 &0 \\
%LDP             & 1    & 1    & 0    & 0    & 1    & 0    & 0    & 1    & 1    & 1    & 1    & 1    & 1\\
%\end{tabular}
%\end{table}
%\end{center}

\begin{center}
\begin{table}[]\label{Table4} \centering
\caption{The predicted values of our method and the LDP method} \label{Table4}
\begin{tabular}{cccccccccccccc}
\hline
Measure  & 1  & 2  & 3  & 4  & 5  & 6  & 7  & 8  & 9  & 10 & 11 & 12 & 13 \\ \hline
Original Value & 0  & 0  & 1  & 0  & 0  & 1  & 0  & 0  & 1  & 0  & 0  & 1  & 0  \\
Ours     & 0  & 0  & 0  & 0  & 1  & 1  & 0  & 1  & 0  & 0  & 0  & 0  & 0  \\
LDP      & 0  & 1  & 0  & 0  & 0  & 1  & 1  & 0  & 1  & 0  & 1  & 0  & 0  \\ \hline
Measure  & 14 & 15 & 16 & 17 & 18 & 19 & 20 & 21 & 22 & 23 & 24 & 25 & 26 \\ \hline
Original Value & 0  & 1  & 0  & 0  & 1  & 0  & 0  & 1  & 0  & 0  & 1  & 0  & 0  \\
Ours     & 0  & 1  & 0  & 1  & 1  & 0  & 0  & 1  & 0  & 0  & 0  & 0  & 0  \\
LDP      & 1  & 1  & 0  & 0  & 1  & 0  & 0  & 1  & 1  & 1  & 1  & 1  & 1  \\ \hline
\end{tabular}
\end{table}
\end{center}

%According to the prediction results in Table \ref{Table4}, we calculate the mean square error (MSE) of the two methods, where the MSE obtained by our method is 0.2355, and the  LDP method's MSE is 0.3846.
According to the prediction results in Table \ref{Table4}, the prediction accuracies of our method and LDP method are 0.73 and 0.62, respectively.
This shows that our method is more accurate. Such finding would indicate that this dataset likely does not follow a sparse model and that previous method was reporting false positives.
In conclusion, our method identifies the 11 most representative significant variables, greatly simplifies the fitting model and presents more accurate results than existing methods. This finding provides a reference for improving the effectiveness of automatic rehabilitation speech assessment tools.

\section{Conclusion}
This paper considers the inference of single  parameter in high-dimensional non-sparse logistic models.
We first find the linearization of the regression model, and then construct the test statistics based on the moment method, which incorporates the null hypothesis.
The proposed procedure is proved to have tight Type I error control even in the dense model.
Our test also has desirable power property. Our test reaches full power quickly when the model is indeed sparse.
It is worth mentioning that the method used in this paper can be extended to sub-Gaussian distribution design and other high dimensional generalized linear models.
For these reasons, our method greatly  complements existing literature.
\backmatter
\section*{Acknowledgments}
This work was supported by  National Social Science Fund project of China [21BTJ045].

\section*{Supplementary information}
\textbf{Supplement of  "Single Parameter Inference of  Non-sparse Logistic Regression Models".}
The detailed proofs about the asymptotic distribution of test statistics  are given. In addition,  we also give detailed proofs of the power property of the test. Technical lemmas are also proved in the supplement.




%\bibliographystyle{plain}
\bibliography{sn-bibliography}% common bib file

\begin{thebibliography}{18}
\providecommand{\natexlab}[1]{#1}
\providecommand{\url}[1]{{#1}}
\providecommand{\urlprefix}{URL }
\providecommand{\doi}[1]{\url{https://doi.org/#1}}
\providecommand{\eprint}[2][]{\url{#2}}
 \bibcommenthead

\bibitem[{Athanasios et~al(2014)Athanasios, Tsanas, Max, Little, Cynthia, Fox,
  Lorraine, and Ramig}]{Athanasios2014Objective}
Athanasios, Tsanas, Max A, et~al (2014) Objective automatic assessment of
  rehabilitative speech treatment in parkinson's disease. IEEE transactions on
  neural systems and rehabilitation engineering: a publication of the IEEE
  Engineering in Medicine and Biology Society

\bibitem[{Bradic et~al(2022)Bradic, Fan, and Zhu}]{2022TESTABILITY}
Bradic J, Fan J, Zhu Y (2022) Testability of high-dimensional linear models
  with nonsparse structures. The Annals of Statistics: An Official Journal of
  the Institute of Mathematical Statistics (2):50

\bibitem[{Cai et~al(2019)Cai, Cai, and Guo}]{2019Optimal}
Cai T, Cai T, Guo Z (2019) Optimal statistical inference for individualized
  treatment effects in high-dimensional models

\bibitem[{Cai et~al(2013)Cai, Liu, and Xia}]{2013Two}
Cai TT, Liu WD, Xia Y (2013) Two-sample covariance matrix testing and support
  recovery in high-dimensional and sparse settings. Journal of the American
  Statistical Association 108(501):265--277

\bibitem[{Dezeure et~al(2017)Dezeure, B\"{u}hlmann, and Zahng}]{Dezeure2017}
Dezeure R, B\"{u}hlmann P, Zahng C (2017) High-dimensional simultaneous
  inference with the bootstrap. TEST 26:685--719

\bibitem[{Van~de Geer et~al(2014)Van~de Geer, B\"{u}hlmann, Ritov, and
  Dezeure}]{2014On}
Van~de Geer S, B\"{u}hlmann P, Ritov Y, et~al (2014) On asymptotically optimal
  confidence regions and tests for high-dimensional models. The Annals of
  Statistics 42(3):1166--1202

\bibitem[{Guo et~al(2021)Guo, Rakshit, Herman, and Chen}]{Guo2021}
Guo Z, Rakshit P, Herman DS, et~al (2021) Inference for the case probability in
  high-dimensional logistic regression. J Mach Learn Res 22(1):1532--4435

\bibitem[{Huang and Zhang(2012)}]{2012Estimation}
Huang J, Zhang CH (2012) Estimation and selection via absolute penalized convex
  minimization and its multistage adaptive applications. Journal of Machine
  Learning Research 13(1):1839--1864

\bibitem[{Lin et~al(2011{\natexlab{a}})Lin, Li, and Zhu}]{LuFeng2011}
Lin L, Li F, Zhu L (2011{\natexlab{a}}) Simulation-based consistent inference
  for biased working model of non-sparse high-dimensional linear regression.
  Journal of Statistical Planning and Inference 141(12):3780--3792

\bibitem[{Lin et~al(2011{\natexlab{b}})Lin, Zhu, and Gai}]{LuLin2011}
Lin L, Zhu L, Gai Y (2011{\natexlab{b}}) Estimation and inference for
  high-dimensional non-sparse models. arXiv:11120712 [statME]

\bibitem[{Ma et~al(2021)Ma, Cai, and Li}]{2021RongMa}
Ma R, Cai TT, Li H (2021) Global and simultaneous hypothesis testing for
  high-dimensional logistic regression models. Journal of the American
  Statistical Association 116(534):984--998

\bibitem[{Ning and Liu(2017)}]{Ning0A2017}
Ning Y, Liu H (2017) A general theory of hypothesis tests and confidence
  regions for sparse high dimensional models. The Annals of Statistics
  45(1):158--195

\bibitem[{Sahand et~al(2012)Sahand, N., Negahban, Pradeep, Ravikumar, Martin,
  J., Wainwright, Bin, and Yu}]{Sahand2012A}
Sahand, N., Negahban, et~al (2012) A unified framework for high-dimensional
  analysis of m-estimators with decomposable regularizers. Statistical Science
  27(4):538--557

\bibitem[{Shi et~al(2021)Shi, Song, Liu, and Li}]{Shi2021}
Shi C, Song R, Liu W, et~al (2021) Statistical inference for high-dimensional
  models via recursive online-score estimation. Joural of the American
  Statistical Association 0(0):1--10

\bibitem[{Sur et~al(2019)Sur, Chen, and Cand\`{e}s}]{Sur2019}
Sur P, Chen Y, Cand\`{e}s EJ (2019) The likelihood ratio test in
  high-dimensional logistic regression is asymptotically a rescaled chi-square.
  Probability Theory and Related Fields 175:487--558

\bibitem[{Zhu and Bradic(2016)}]{2016Two}
Zhu YC, Bradic J (2016) Two-sample testing in non-sparse high-dimensional
  linear models. arXiv preprint arXiv p 1610.04580

\bibitem[{Zhu and Bradic(2017)}]{2016Linear}
Zhu YC, Bradic J (2017) Linear hypothesis testing in dense high-dimensional
  linear models. Journal of the American Statal Association 113(524):1583--1600

\bibitem[{Zhu and Bradic(2018)}]{2018Significance}
Zhu YC, Bradic J (2018) Significance testing in non-sparse high-dimensional
  linear models. Electronic Journal of Statistics 12(2):3312--3364

\end{thebibliography}


\begin{thebibliography}{2}
\expandafter\ifx\csname natexlab\endcsname\relax\def\natexlab#1{#1}\fi
\providecommand{\url}[1]{\texttt{#1}}
\providecommand{\href}[2]{#2}
\providecommand{\path}[1]{#1}
\providecommand{\DOIprefix}{doi:}
\providecommand{\ArXivprefix}{arXiv:}
\providecommand{\URLprefix}{URL: }
\providecommand{\Pubmedprefix}{pmid:}
\providecommand{\doi}[1]{\href{http://dx.doi.org/#1}{\path{#1}}}
\providecommand{\Pubmed}[1]{\href{pmid:#1}{\path{#1}}}
\providecommand{\bibinfo}[2]{#2}
\ifx\xfnm\relax \def\xfnm[#1]{\unskip,\space#1}\fi
%Type = Article
\bibitem[{Vershynin(2010)}]{2010Introduction}
\bibinfo{author}{Vershynin, R.}, \bibinfo{year}{2010}.
\newblock \bibinfo{title}{Introduction to the non-asymptotic analysis of random
  matrices}.
\newblock \bibinfo{journal}{arXiv preprint arXiv} , \bibinfo{pages}{1011.3027}.
%Type = Article
\bibitem[{Zhu and Bradic(2016)}]{2016Two}
\bibinfo{author}{Zhu, Y.C.}, \bibinfo{author}{Bradic, J.},
  \bibinfo{year}{2016}.
\newblock \bibinfo{title}{Two-sample testing in non-sparse high-dimensional
  linear models}.
\newblock \bibinfo{journal}{arXiv preprint arXiv} ,
  \bibinfo{pages}{1610.04580}.

\end{thebibliography}

\end{document}

\begin{frontmatter}



\title{ Supplement of "Single Parameter Inference of  Non-sparse Logistic Regression Models" \tnoteref{t1,t2}}
\tnotetext[t1]{This work was supported by the National Social Science Fund project of China [21BTJ045].}
%%             country={}}\corref{cor1

\author[1]{Yanmei Shi\fnref{fn1}}
\ead{2020020263@qdu.edu.cn}

%\author[1]{Zhiruo Li}
%\ead{2020025413@qdu.edu.cn}

\author[1]{Qi Zhang \corref{cor1}}
\ead{qizhang@qdu.edu.cn}

 \cortext[cor1]{Corresponding author}
 \fntext[fn1]{The first author.}
 %\fntext[fn2]{Another author footnote.}
% \fntext[fn3]{Yet another author footnote.}

%\affiliation[1]{organization={Elsevier B.V.},
%                 addressline={Radarweg 29},
%                 postcode={1043 NX},
%                 city={Amsterdam},
%                 country={The Netherlands}}

 \affiliation[1]{organization={Institute of Mathematics and Statistics},
                 addressline={Qingdao University},
                 postcode={266071},
                 city={Qingdao},
                 country={China}}
 %\affiliation[3]{organization={Sayahna Foundation},
%                 addressline={JWRA 34, Jagathy},
%                 postcode={695014},
%                 city={Trivandrum},
%                 country={China}}
%\affiliation{organization={},%Department and Organization

\begin{abstract}
This note summarizes the supplementary materials to the paper "Single Parameter Inference of  Non-sparse Logistic Regression Models". We present the detailed proofs of  size and power properties.
 We also prove the  technical lemmas.
\end{abstract}



%\begin{keyword}
%Logistic models \sep Non-sparse \sep Single parameter hypothesis test \sep Moment condition
%%% keywords here, in the form: keyword \sep keyword
%\MSC 62F03  \sep 62F35 \sep 62J15
%\end{keyword}

\end{frontmatter}

\section{Proof of Theorem 1}

First we consider a map $\mathcal{B}$ that satisfies: \\
($i$) For any $\sigma$, $\mathcal{B} \subseteq \mathbb{R}^{p}$; \\
($ii$) Monotonicity: for $ \sigma_{1}\leq\sigma_{2}$, $\mathcal{B}(\sigma_{1})\subseteq\mathcal{B}(\sigma_{2})$.
\begin{lemma} \label{lemma 4.2}
Let Assumption 3.1 be established, then \\
($i$) $\|W(\tilde{\pi}-\pi)\|_{2}^{2}=O_{p}(s_{\pi}logp)$ \notag \\
and $ \|\tilde{\pi}-\pi\|_{1}=O_{p}(s_{\pi}\sqrt{n^{-1}logp})$ \notag; \\
($ii$) $n^{-\frac{1}{2}}(\pi-\tilde{\pi})^{T}W^{T}(V-W\tilde{\theta})\hat{\sigma}_{e}^{-1}=O_{p}(s_{\pi}n^{-\frac{1}{2}}logp)$.
\end{lemma}
%\noindent \emph{\textbf{Proof of Lemma \ref{lemma 4.2}}}
\begin{proof}
The constraints in the definition of $\tilde{\pi}$ are equivalent to
\begin{align} \tag{1.1} \label{constraint2}
\begin{cases}
\|n^{-1}W^{T}(Z-W\pi)\|_{\infty}\leq \eta\sigma_{u}, \\
\|Z-W\pi\|_{2}^{2}\geq (\frac{\rho_{0}^{2}}{2})n\sigma_{u}^{2}/2.
\end{cases}
\end{align}
Applying Theorem 4 in \cite{2016Two} with $( H ,G, b_{*}, \varepsilon)=(Z,W,\pi,U)$, we can give (i) without proof.
The constraints  in the definition of $\tilde{\theta}$ are equivalent to
\begin{align} \tag{1.2} \label{constraint1}
\begin{cases}
\|n^{-1}W^{T}(V-W\theta)\|_{\infty}\leq \eta\sigma_{e}, \\
\|V-W\theta\|_{2}^{2}\geq  (\frac{\rho_{0}^{2}}{2})n\sigma_{e}^{2}/2.
\end{cases}
\end{align}
%
%Since the constraints in \eqref{constraint1}, we have
%\begin{align}
%\|n^{-1}W^{T}(V-W\tilde{\theta})\|_{\infty}\leq \eta\tilde{\sigma_{e}} \tag{1.3} \label{A*}.
%\end{align}
Moreover,
\begin{align}
n^{-\frac{1}{2}}(\pi-\tilde{\pi})^{T}W^{T}(V-W\tilde{\theta})\hat{\sigma}_{e}^{-1}
\leq\vert n^{-\frac{1}{2}}(\pi-\tilde{\pi})^{T}W^{T}(V-W\tilde{\theta})\hat{\sigma}_{e}^{-1}\vert \tag{1.3} \label{A**}.
\end{align}
According to Hold's inequality, we have
\begin{align}
\eqref{A**}\leq \sqrt{n}\|\pi-\tilde{\pi}\|_{1}\|n^{-1}W^{T}(V-W\tilde{\theta})\|_{\infty}\hat{\sigma}_{e}^{-1}
\tag{1.4} \label{A***}.
\end{align}
By \eqref{constraint1} and (i), we get
\begin{align}
\eqref{A***}\leq \sqrt{n}O_{p}(s_{\pi}\sqrt{n^{-1}logp})\eta\tilde{\sigma_{e}}\hat{\sigma}_{e}^{-1} \tag{1.5}  \label{A****}.
\end{align}
According to the constraint in \eqref{constraint1}: $\|V-W\tilde{\theta}\|_{2}^{2}\geq n(\frac{\rho_{0}^{2}}{2})\tilde{\sigma}_{e}^{2}/2$ \ and  $\hat{\sigma}_{e}=\|V-W\tilde{\theta}\|_{2}/\sqrt{n}$, we get
\begin{align}
\eqref{A****}\leq& \sqrt{n}O_{p}(s_{\pi}\sqrt{n^{-1}logp})\eta\frac{2}{\sqrt{n}\rho_{0}}
\|V-W\tilde{\theta}\|_{2}\frac{\sqrt{n}}{\|V-W\tilde{\theta}\|_{2}} \notag \\
=&\frac{\sqrt{2n}}{\rho_{0}}O_{p}(s_{\pi}\sqrt{n^{-1}logp})\eta \tag{1.6} \label{A*****}.
\end{align}
By $\eta\asymp \sqrt{n^{-1}logp}$, then
\begin{align}
\eqref{A*****}=\frac{\sqrt{2n}}{\rho_{0}}O_{p}(s_{\pi}n^{-1}logp)=O_{p}(s_{\pi}logp/\sqrt{n}) \notag.
\end{align}
Therefore,
\begin{align}
n^{-\frac{1}{2}}(\pi-\tilde{\pi})^{T}W^{T}(V-W\tilde{\theta})\hat{\sigma}_{e}^{-1}=O_{p}(s_{\pi}n^{-\frac{1}{2}}logp) \notag.
\end{align}
The proof is complete.
\end{proof}
\begin{lemma} \label{lemma 4.3}
Let Assumption 3.1 be established, then
\begin{equation}
\vert\hat{Q}-E(u_{1}u_{1}^{T})\vert=O_{p}((s_{\pi}n^{-1}logp)\vee \notag \sqrt{n^{-1}logp }).
\end{equation}
\end{lemma}
%\noindent \emph{\textbf{Proof of Lemma \ref{lemmaQ_hat}}}
\begin{proof}
Observe that
\begin{align}
&\vert\hat{Q}-Eu_{1}^{2}\vert \notag\\
=&\vert n^{-1}\sum\limits_{i=1}^n(\hat{u}_{i}^{2}-Eu_{1}^{2})\vert \notag \\
=&\vert n^{-1}\sum\limits_{i=1}^n(\hat{u}_{i}^{2}-u_{i}^{2}+u_{i}^{2}-Eu_{1}^{2})\vert \notag \\
\leq& \vert n^{-1}\sum\limits_{i=1}^n(\hat{u}_{i}^{2}-u_{i}^{2})\vert+\vert n^{-1}\sum\limits_{i=1}^n(u_{i}^{2}-Eu_{1}^{2})\vert \tag {1.7} \label{A.1}
\end{align}
We constrain the two terms in  \eqref{A.1} respectively. Since $u_{1}$ is a Gaussian term with bounded variance,   $u_{i}^{2}$ has sub-exponential norms bounded above by some constant $K>0$. Therefore, by Proposition 5.16 of \cite{2010Introduction},  for any $ t >0 $, we have
\begin{align}
&P(\mid n^{-1}\sum\limits_{i=1}^n(u_{i}^{2}-Eu_{1}^{2})\mid>t\sqrt{n^{-1}logp})
%\leq& \sum\limits_{j_{1}=1}^p\sum\limits_{j_{2}=1}^pP(\vert \sum\limits_{i=1}^nu_{i,j_{1}}u_{i,j_{2}}- Eu_{1,j_{1}}u_{2,j_{2}}\vert >t\sqrt{nlogp}) \notag\\
\leq& 2exp[-c\min(\frac{t^{2}nlogp}{K^{2}n},\frac{t\sqrt{nlogp}}{K})]\rightarrow 0 \notag.
\end{align}
where $c>0$ is a universal constant. Therefore, by taking $t=4K^{2}/c$, we have
\begin{align}
\vert n^{-1}\sum\limits_{i=1}^n(u_{i}^{2}-Eu_{1}^{2})\vert = O_{p}(\sqrt{n^{-1}logp}). \tag{1.8} \label{A.2}
\end{align}
In addition, notice that
\begin{align}
&\hat{u}_{i}^{2}-u_{i}^{2} \notag \\
=&[z_{i}-\tilde{\pi}^{T}w_{i}]^{2}-u_{i}^{2} \notag \\
=&[\pi^{T}w_{i}+u_{i}-\tilde{\pi}^{T}w_{i}]^{2}-u_{i}^{2} \notag \\
=&[(\pi-\tilde{\pi})^{T}w_{i}+u_{i}]^{2}-u_{i}^{2} \notag \\
=&((\pi-\tilde{\pi})^{T}w_{i})^{2}+2(\pi-\tilde{\pi})^{T}w_{i}u_{i}
+u_{i}^{2} -u_{i}^{2} \notag \\
=&((\pi-\tilde{\pi})^{T}w_{i})^{2}+2(\pi-\tilde{\pi})^{T}w_{i}u_{i} \notag.
\end{align}
Therefore,
\begin{align}
&\vert n^{-1}\sum\limits_{i=1}^n(\hat{u}_{i}^{2}-u_{i}^{2})\vert \notag \\
\leq& \vert n^{-1}\sum\limits_{i=1}^n((\pi-\tilde{\pi})^{T}w_{i})^{2}\vert
+2\vert n^{-1}\sum\limits_{i=1}^n(\pi-\tilde{\pi})^{T}w_{i}u_{i}\vert \notag \\
=&\|n^{-\frac{1}{2}}W(\tilde{\pi}-\pi)\|_{2}^{2}
+2\vert n^{-1}\sum\limits_{i=1}^n(\pi-\tilde{\pi})^{T}w_{i}u_{i}\vert \tag{1.9} \label{A.3}.
\end{align}
For $\|n^{-\frac{1}{2}}W(\tilde{\pi}-\pi)\|_{2}^{2}$, by (i) in Lemma  \ref{lemma 4.2}, we have
%\begin{align}
%\max\limits_{j \in G}\|W(\tilde{\pi}_{j}-\pi_{j})\|_{2}^{2}=O_{p}(s_{\pi}logp) \notag
%\end{align}
%Therefore,
\begin{align}
\|n^{-\frac{1}{2}}W(\tilde{\pi}-\pi)\|_{2}^{2}=O_{p}(n^{-1}s_{\pi}logp) \tag{1.10} \label{A.4}.
\end{align}
For $2\vert n^{-1}\sum\limits_{i=1}^n(\pi-\tilde{\pi})^{T}w_{i}u_{i}\vert $, according to Hold's inequality, we have
\begin{align}
2\vert n^{-1}\sum\limits_{i=1}^n(\pi-\tilde{\pi})^{T}w_{i}u_{i}\vert \leq 2\|n^{-1}\sum\limits_{i=1}^n u_{i}w_{i}^{T}\|_{\infty}\|\pi-\tilde{\pi}\|_{1} \notag.
\end{align}
By (i) in Lemma  \ref{lemma 4.2}: $\|\tilde{\pi}-\pi\|_{1}=O_{p}(s_{\pi}\sqrt{n^{-1}logp})$, we have
\begin{align}
2\|n^{-1}\sum\limits_{i=1}^n u_{i}w_{i}^{T}\|_{\infty}\|\pi-\tilde{\pi}\|_{1} =\|n^{-1}\sum\limits_{i=1}^n u_{i}w_{i}^{T}\|_{\infty}O_{p}(s_{\pi}\sqrt{n^{-1}logp}) \notag.
\end{align}
Therefore,
\begin{align}
2\vert n^{-1}\sum\limits_{i=1}^n(\pi-\tilde{\pi})^{T}w_{i}u_{i}\vert \leq \|n^{-1}\sum\limits_{i=1}^n u_{i}w_{i}^{T}\|_{\infty}O_{p}(s_{\pi}\sqrt{n^{-1}logp}) \notag.
\end{align}
According to the sub-Gaussian  properties of $u_{i}$ and $w_{i}$,  the terms of $u_{i}w_{i}^{T}$ have bounded exponential norm.  Proposition 5.16 of \cite{2010Introduction} and the union bound imply that
\begin{align}
P(\|n^{-1}\sum\limits_{i=1}^n u_{i}w_{i}^{T}\|_{\infty}\geq t\sqrt{n^{-1}logp})\leq 2(p-1)exp[-c\min(\frac{t^{2}nlogp}{K^{2}n},\frac{t\sqrt{nlogp}}{K})]\rightarrow 0 \notag.
\end{align}
Therefore, by taking $t=4K^{2}/c$, we have
\begin{align}
 \|n^{-1}\sum\limits_{i=1}^n u_{i}w_{i}^{T}\|_{\infty}=O_{p}(\sqrt{n^{-1}logp}) \notag.
\end{align}
Therefore,
\begin{align}
2\vert n^{-1}\sum\limits_{i=1}^n(\pi-\tilde{\pi})^{T}w_{i}u_{i}\vert = O_{p}(\sqrt{n^{-1}logp})O_{p}(s_{\pi}\sqrt{n^{-1}logp}) = O_{p}(s_{\pi}n^{-1}logp) \tag{1.11} \label{A.5}.
\end{align}
 %Since $s_{\pi}\asymp \sqrt{\frac{n}{log^{3}p}}$.
 By \eqref{A.4} and \eqref{A.5}, we have
\begin{align}
\vert n^{-1}\sum\limits_{i=1}^n(\hat{u}_{i}^{2}-u_{i}^{2})\vert=O_{p}(n^{-1}s_{\pi}logp) \tag{1.12} \label{A.6}.
\end{align}
 Combining \eqref{A.1}, \eqref{A.2} and \eqref{A.6}, we can get
 \begin{align}
 \vert\hat{Q}-Eu_{1}^{2}\vert=O_{p}((s_{\pi}n^{-1}log p)\vee \sqrt{n^{-1}logp} ) \notag.
 \end{align}
 The proof is complete.
\end{proof}
In addition, since $s_{\pi}=o(\sqrt{n/log^{3}p})$, so
\begin{align}
\vert\hat{Q}-Eu_{1}^{2}\vert=O_{p}(\sqrt{n^{-1}log^{-1}p}\vee \sqrt{n^{-1}logp})=O_{p}(\sqrt{n^{-1}logp}) \notag.
\end{align}
\noindent \emph{\textbf{Proof of Theorem 1.}}
\begin{proof}
Consider the test statistic $T_{n}=n^{-\frac{1}{2}}\hat{\sigma}_{e}^{-1}(Z-W\tilde{\pi})^{T}(V-W\tilde{\theta})
$ under the null hypothesis $H_{0}$,
\begin{eqnarray}
T_{n}
&\leq&\vert n^{-\frac{1}{2}}(W(\pi-\tilde{\pi})+U)^{T}(V-W\tilde{\theta})\hat{\sigma}_{e}^{-1}\vert
\nonumber  \\
&\leq&\vert n^{-\frac{1}{2}}\hat{\sigma}_{e}^{-1}(\pi-\tilde{\pi})^{T}W^{T}(V-W\tilde{\theta})\vert+\vert n^{-\frac{1}{2}}\hat{\sigma}_{e}^{-1}U^{T}(V-W\tilde{\theta})\vert \nonumber \\
&=&\vert\triangle_{n}\vert+\vert n^{-\frac{1}{2}}\hat{\sigma}_{e}^{-1}U^{T}(V-W\tilde{\theta})\vert
\nonumber,
 \end{eqnarray}
where $\triangle_{n}=n^{-\frac{1}{2}}\hat{\sigma}_{e}^{-1}(\pi-\tilde{\pi})^{T}W^{T}(V-W\tilde{\theta})$.
%where $\Gamma(x;Q):=P(\|\xi\|_{\infty}\leq x),\ \xi \sim N(0, Q)$.
According to Lemma \ref{lemma 4.2}, we have
\begin{align}
\Delta_{n}= n^{-\frac{1}{2}}\hat{\sigma}_{e}^{-1}(\pi-\tilde{\pi})^{T}W^{T}(V-W\tilde{\theta}) =O_{p}(s_{\pi}n^{-\frac{1}{2}}logp)
=o_{p}(\frac{1}{\sqrt{logp}}) \notag .
 \end{align}
 Therefore, the behavior of the test statistic is completely determined by $\vert n^{-\frac{1}{2}}\hat{\sigma}_{e}^{-1}U^{T}(V-W\tilde{\theta})\vert$.
 Let $\mathcal{F}_{n}$ be the $\sigma$-algebra generated by $\varepsilon$ and $W$, under the null hypothesis $H_{0}: \beta_{*}=\beta_{0}$,
\begin{eqnarray}
V-W\tilde{\theta}
=Y-Z\beta_{0}-W\tilde{\theta}
=Z(\beta_{*}-\beta_{0})+W(\theta_{*}-\tilde{\theta})+\varepsilon
=W(\theta_{*}-\tilde{\theta})+\varepsilon \nonumber.
\end{eqnarray}

 Note that the Modified Dantzig Selector estimator $\tilde{\theta}$ is a function of $(V,W) $.
  Since $\hat{\sigma}_{e}=\frac{\|V-W\tilde{\theta}\|_{2}}{\sqrt{n}}$,  $\hat{\sigma}_{e}$ is also a function of $V$ and $W$. And $U$ is independent of  $V-W\tilde{\theta}$.
Under the assumption of a Gaussian distribution of $U$, for $n^{-\frac{1}{2}}\hat{\sigma}_{e}^{-1}U^{T}(V-W\tilde{\theta})=n^{-\frac{1}{2}}\sum\limits_{i=1}^{n}u_{i}\hat{e}_{i}\hat{\sigma}_{e}^{-1}$ and the given $\mathcal{F}_{n}$, we have
\begin{align}
E[n^{-\frac{1}{2}}\sum\limits_{i=1}^{n}u_{i}\hat{e}_{i}\hat{\sigma}_{e}^{-1}\vert\mathcal{F}_{n}]=&\frac{1}{\|\hat{e}\|_{2}}\sum\limits_{i=1}^{n}\hat{e}_{i}E(u_{i})=0 \notag, \\
Var[n^{-\frac{1}{2}}\sum\limits_{i=1}^{n}u_{i}\hat{e}_{i}\hat{\sigma}_{e}^{-1}\vert\mathcal{F}_{n}]
=&n^{-1}\hat{\sigma}_{e}^{-2}\sum\limits_{i=1}^{n}\hat{e}_{i}^{2}Var(u_{i})
=E(u_{1}^{2}) \notag,
\end{align}
Let $Q=E(u_{1}^{2})$, then
\begin{align}
n^{-\frac{1}{2}}\hat{\sigma}_{e}^{-1}U^{T}(V-W\tilde{\theta})\sim N(0,Q)
\notag.
\end{align}
In other words,
\begin{align}
p(n^{-\frac{1}{2}}\hat{\sigma}_{e}^{-1}U^{T}(V-W\tilde{\theta}))\leq x)=\Phi(x;Q)
\notag,
\end{align}
%where $\Gamma(x;A)=\vert N(0,A)\vert$.
By Lemma \ref{lemma 4.3}, we have
\begin{eqnarray}
\vert\hat{Q}-Q\vert&=&O_{p}((s_{\pi}n^{-1}logp)\vee \sqrt{n^{-1}logp})
=o_{p}(\sqrt{n^{-1}logp}) \nonumber.
\end{eqnarray}
The above conclusions  make the assumptions in Theorem 3 with $(G_{n},\Psi_{i},\Delta_{n})=(T_{n}, u_{i}\hat{e}_{i}\hat{\sigma}_{e}^{-1}, \Delta_{n} )$ in \cite{2016Two} all satisfied. Then
\begin{align}
P( T_{n}\leq x ) = N(0, \hat{Q})+o_{p}(1), \ {\forall} x \geq 0. \notag
\end{align}
Finally, we get
\begin{align}
P(T_{n}>\Phi^{-1}(1-\alpha/2;\hat{Q}))\rightarrow \alpha \notag.
\end{align}

The proof is complete.
\end{proof}

\section{Proof of Theorem 2}
\begin{eqnarray}
V=Y-Z\beta_{0}=Z\beta_{*}+W\theta+\varepsilon-Z\beta_{0}
=W(\theta_{*}+\pi(\beta_{*}-\beta_{0}))+U(\beta_{*}-\beta_{0})+\varepsilon \nonumber.
\end{eqnarray}
%\begin{lemma} \label{lemma 4.5}
%Let $\{x_{i}\}_{i=1}^{n}$ and $\{h_{i}\}_{i=1}^{n}$ be sequences of two random variables of dimension $\mathbb{R}^{L}$ that are independent across $i$. Suppose that $\{x_{i}\}_{i=1}^{n}$ and $\{h_{i}\}_{i=1}^{n}$ are also independent and that there exist constants $K_{1}, K_{2} > 0$ such that ${\forall}$ $ 1\leq l \leq L$, the sub-Gaussian norm of $x_{i,l}$ is bounded above by $K_{1}$ and $P(\max\limits_{1\leq l\leq L}n^{-1}\sum_{i=1}^{n}h_{i,l}^{2}>K_{2})\rightarrow 0 $ .
%Then, if $L \rightarrow \infty$, there exists a constant $K>0$ depending only on $K_{1}, K_{2}$ such that
%\begin{align}
%P(\max\limits_{1\leq l\leq L}\vert n^{-\frac{1}{2}}\sum\limits_{i=1}\limits^{n}x_{i,l}h_{i,l}\vert>K\sqrt{log L})\rightarrow 0 \notag
%\end{align}
%\end{lemma}
\begin{lemma} \label{lemma 4.6}
Let $U_{*}=U(\beta_{*}-\beta_{0})+\varepsilon$ and Assumption 3.2 be hold, then
\begin{align}
\frac{3}{4}\sigma_{*}^{2}\leq n^{-1}\|U_{*}\|_{2}^{2}\leq 2\sigma_{*} \notag,\\
\|n^{-1}W^{T}U_{*}\|_{\infty}\leq \eta\sigma_{*} \notag.
\end{align}
where $\sigma_{*}=\sqrt{(\beta_{*}-\beta_{0})^{2}\sigma_{u}^{2}+\sigma_{\varepsilon}^{2}}$.
 %with $\sigma_{u}^{2}=n^{-1}\sum\limits_{i=1}^{n}u_{i}^{2}$.\\
\end{lemma}
%\noindent \emph{\textbf{Proof of Lemma \ref{lemmaW U}:}}
\begin{proof}
Notice that the entries of $\frac{U_{*}}{\sigma_{*}}=\frac{U(\beta_{*}-\beta_{0})+\varepsilon}{\sqrt{(\beta_{*}-\beta_{0})^{2}\sigma_{u}^{2}+\sigma_{\varepsilon}^{2}}}$ are $i.i.d $ $ \mathcal{N}(0,1)$ random variables that are independent of $W$. We apply Lemma 19 in \cite{2016Two}  with $L=p-1$, $x_{i,l}=w_{i,l}$ and $h_{i}=((u_{i}(\gamma-\beta_{G}^{0})+\varepsilon_{i})/\sigma_{*},0,...,0)$ for $1\leq l \leq L $, then
\begin{align}
P(\max\limits_{1\leq l\leq L}\vert n^{-\frac{1}{2}}\sum\limits_{i=1}^{n}x_{i,l}h_{i,l}\vert> \eta\sigma_{*})\rightarrow 0 \notag,
%P(\|n^{-1}W^{T}U_{\gamma}\|_{\infty}>\eta\sigma_{*})\rightarrow 0 \notag.
\end{align}
which suggests that $P(\|n^{-1}W^{T}U_{*}\|_{\infty}>\eta\sigma_{*})\rightarrow 0$.
Notice that $n^{-1}\|U_{*}\|_{2}^{2}\sigma_{*}^{-2}$ is the average of independent $\mathcal{X}^{2}(1)$ random variables. By the law of large numbers,  $n^{-1}\|U_{*}\|_{2}^{2}\sigma_{*}^{-1}=1+o_{p}(1)$. Therefore,
\begin{align}
\frac{3}{4}\sigma_{*}^{2}\leq n^{-1}\|U_{*}\|_{2}^{2}\leq 2\sigma_{*} \notag.
\end{align}

The proof is complete.
\end{proof}
\noindent \emph{\textbf{Proof of Theorem 2}}.
\begin{proof}
We apply Theorem 4 in \cite{2016Two} with $( H ,G, b_{*},
\varepsilon)=(V,W,\theta_{*}+\pi(\beta_{*}-\beta_{0}),U(\beta_{*}-\beta_{0})+\varepsilon)
$ and $\mathcal{B}(\sigma)=\mathbb{R}^{p}$, ${\forall}$ $\sigma \geq 0$. Notice that the restricted eigenvalue condition holds for some
constant $\kappa >0$ with probability approaching 1. Moreover,
\begin{align}
\|\theta_{*}+\pi(\beta_{*}-\beta_{0})\|_{0} \leq& \|\theta_{*}\|_{0}+\|\pi(\beta_{*}-\beta_{0})\|_{0}
 \notag \\
\leq& \|\theta_{*}\|_{0}+ \|\pi\|_{0}\|(\beta_{*}-\beta_{0})\|_{0} \notag \\
=& \|\theta_{*}\|_{0}+s_{\pi} \|(\beta_{*}-\beta_{0})\|_{0}= s_{*} \notag.
\end{align}
By the  $\|\theta_{*}\|_{0}=o(\sqrt{n}/logp)$ in Assumption 3.2, we have
$s_{*}=\|\theta_{*}\|_{0}+s_{\pi} \|(\beta_{*}-\beta_{0})\|_{0}= o(\sqrt{n}/logp)$. Therefore,
\begin{align}
\|\theta_{*}+\pi(\beta_{*}-\beta_{0})\|_{0}= o(\sqrt{n}/logp) \notag,
\end{align}
and
\begin{align}
28\eta\sqrt{\|\theta_{*}+\pi(\beta_{*}-\beta_{0})\|_{0}/\kappa} \leq 28\eta\sqrt{o(\sqrt{n}/logp)/\kappa}\rightarrow 0 \notag.
\end{align}
The above conclusions and Lemma \ref{lemma 4.6} make the assumptions in Theorem 4 in \cite{2016Two} hold, then

\begin{align}\tag{1.13} \label{A.7}
\begin{cases}
&\sigma_{*}/\sqrt{2} \leq \hat{\sigma}_{e} \leq 2\sigma_{*}  \ \ \  (1) \\
&\|\pi(\beta_{*}-\beta_{0})+\theta_{*}-\tilde{\theta}\|_{1}\hat{\sigma}_{e}^{-1}\leq 16s_{*}\eta\sigma_{*}\hat{\sigma}_{e}^{-1}/\kappa \leq 16\sqrt{2}s_{*}\eta/\kappa \ \ \ (2) \\
&\|n^{-1}W^{T}(V-W\tilde{\theta})\|_{\infty}\hat{\sigma}_{e}^{-1} \leq 3\sqrt{2}\eta \ \ \ (3).
\end{cases}
\end{align}
where, $\hat{\sigma}_{e}=n^{-\frac{1}{2}}\|V-W\tilde{\theta}\|_{2}$. \\
%Let $j_{*}\in G$ satisfies that $\|\Sigma_{U}(\gamma-\beta_{G}^{0})\|_{\infty}=\vert e_{j_{*}}^{T}\Sigma_{U}(\gamma-\beta_{G}^{0})\vert$, where $e_{j_{*}} \in \mathbb{R}^{p} $ is the $j_{*}$-th column of $I_{p}$. \\
{\bf{Step 1: derive the behavior of the test statistic. }}
\begin{align}
T_{n}=& n^{-\frac{1}{2}}(Z-W\tilde{\pi})^{T}(V-W\tilde{\theta})\hat{\sigma}_{e}^{-1} \notag \\
%\geq& \vert n^{-\frac{1}{2}}(z_{j*}-W\tilde{\pi}_{j*})^{T}(V-W\tilde{\theta})\hat{\sigma}_{e}^{-1}\vert  \notag \\
=&\vert n^{-\frac{1}{2}}(W(\pi-\tilde{\pi})+U)^{T}(W(\pi(\beta_{*}-\beta_{0})+(\theta_{*}-\tilde{\theta}))+U(\beta_{*}-\beta_{0})+\varepsilon)\hat{\sigma}_{e}^{-1}\vert \notag \\
\geq& \vert n^{-\frac{1}{2}}U^{T}U(\beta_{*}-\beta_{0})\hat{\sigma}_{e}^{-1}\vert - \vert n^{-\frac{1}{2}}U^{T}W(\pi(\beta_{*}-\beta_{0})+(\theta_{*}-\tilde{\theta}))\hat{\sigma}_{e}^{-1}\vert  - \vert n^{-\frac{1}{2}}U^{T}\varepsilon\hat{\sigma}_{e}^{-1}\vert  \notag \\
-&\vert n^{-\frac{1}{2}}(\pi-\tilde{\pi})^{T}W^{T}(W(\pi(\beta_{*}-\beta_{0})+(\theta_{*}-\tilde{\theta}))+U(\beta_{*}-\beta_{0})+\varepsilon)\hat{\sigma}_{e}^{-1} \vert
\notag \\
=&J_{1}-J_{2}-J_{3}-J_{4} \notag.
\end{align}
where $J_{1}=\vert n^{-\frac{1}{2}}U^{T}U(\beta_{*}-\beta_{0})\hat{\sigma}_{e}^{-1}\vert $, $J_{2}=\vert n^{-\frac{1}{2}}U^{T}W(\pi(\beta_{*}-\beta_{0})+(\theta_{*}-\tilde{\theta}))\hat{\sigma}_{e}^{-1}\vert$, $J_{3}=\vert n^{-\frac{1}{2}}U^{T}\varepsilon\hat{\sigma}_{e}^{-1}\vert$ and $J_{4}=\vert n^{-\frac{1}{2}}(\pi-\tilde{\pi})^{T}W^{T}(W(\pi(\beta_{*}-\beta_{0})+(\theta_{*}-\tilde{\theta}))+U(\beta_{*}-\beta_{0})+\varepsilon)\hat{\sigma}_{e}^{-1} \vert$.

We first analyze $J_{4}$,
\begin{align}
J_{4}=&\vert n^{-\frac{1}{2}}(\pi-\tilde{\pi})^{T}W^{T}(W(\pi(\beta_{*}-\beta_{0})+(\theta_{*}-\tilde{\theta}))+U(\beta_{*}-\beta_{0})+\varepsilon)\hat{\sigma}_{e}^{-1} \vert  \notag \\
=&\vert n^{-\frac{1}{2}}(\pi-\tilde{\pi})^{T}W^{T}(V-W\tilde{\theta})\hat{\sigma}_{e}^{-1} \vert  \notag \\
\leq & \sqrt{n}\|\pi-\tilde{\pi}\|_{1}\|n^{-1}W^{T}(V-W\tilde{\theta})\|_{\infty}\hat{\sigma}_{e}^{-1} \notag \\
\leq& \sqrt{n}O_{p}(s_{\pi}\sqrt{n^{-1}logp})3\sqrt{2}\eta \notag \\
=&o_{p}(1/\sqrt{logp}) \tag{1.14} \label{A.8}.
\end{align}
where the first inequality is established according to the Hold's inequality;  the second inequality is established according to Lemma \ref{lemma 4.2} and (3) in \eqref{A.7}. The last equation holds according to $s_{\pi}=o(\sqrt{n/log^{3}p})$.

Since $u_{i}$ has a sub-Gaussian norm, the law of large numbers implies that
\begin{align}
n^{-1}\sum\limits_{i=1}^{n}u_{i}^{2}=\sigma_{u}^{2}+o_{p}(1) \notag.
\end{align}
Therefore, for any constant $C_{1}$ that satisfies that $C_{1}\geq\sigma_{u}^{2}$, we have
\begin{align}
P(n^{-1}\sum\limits_{i=1}^{n}u_{i}^{2}>C_{1}+1)\rightarrow 0 \notag.
\end{align}
According to the sub-Gaussian property of $w_{i}$, applying the Lemma 19 in \cite{2016Two} with $x_{i,j}=w_{i,j}$ and $h_{i}=(u_{i},0,...,0)$, for $1\leq j \leq p-1$, we have
\begin{align}
\|n^{-\frac{1}{2}}U^{T}W\|_{\infty}=\max\limits_{1\leq j \leq p-1}\vert n^{-\frac{1}{2}}\sum \limits_{i=1}^{n}w_{i,j}u_{i,j}\vert =O_{p}(\sqrt{log p}) \notag.
\end{align}
By Hold's inequality, we have
\begin{align}
J_{2}=&\vert n^{-\frac{1}{2}}U^{T}W(\pi(\beta_{*}-\beta_{0})+(\theta_{*}-\tilde{\theta}))\hat{\sigma}_{e}^{-1}\vert \notag \\
\leq & \|n^{-\frac{1}{2}}U^{T}W\|_{\infty}\|\pi(\beta_{*}-\beta_{0})+(\theta_{*}-\tilde{\theta})\|_{1}\hat{\sigma}_{e}^{-1} \notag \\
=&O_{p}(\sqrt{logp})\|\pi(\beta_{*}-\beta_{0})+(\theta_{*}-\tilde{\theta})\|_{1}\hat{\sigma}_{e}^{-1} \notag.
\end{align}
According to (2) in \eqref{A.7}: $\|\pi(\beta_{*}-\beta_{0})+\theta_{*}-\tilde{\theta}\|_{1}\hat{\sigma}_{e}^{-1}\leq 16\sqrt{2}s_{*}\eta/\kappa$, then
\begin{align}
O_{p}(\sqrt{logp})\|\pi(\beta_{*}-\beta_{0})+(\theta_{*}-\tilde{\theta})\|_{1}\hat{\sigma}_{e}^{-1} \leq O_{p}(\sqrt{logp})16\sqrt{2}s_{*}\eta/\kappa \notag.
\end{align}
And since $s_{*}=\|\theta_{*}\|_{0}+s_{\pi} \|(\beta_{*}-\beta_{0})\|_{0}= o(\sqrt{n}/logp)$, we get
\begin{align}
O_{p}(\sqrt{logp})16\sqrt{2}s_{*}\eta/\kappa=o_{p}(\sqrt{logp}\frac{\sqrt{n}}{logp}\frac{\sqrt{logp}}{\sqrt{n}})=o_{p}(1) \notag.
\end{align}
Therefore,
\begin{align}
J_{2}=o_{p}(1)  \tag{1.15} \label{A.9}.
\end{align}

Notice that $E\vert u_{i}\varepsilon_{i}\vert^{2+\delta}=E\vert u_{i}\vert ^{2+\delta}E\vert \varepsilon_{i}\vert^{2+\delta}$, which is bounded by the constant $C_{2}>0$, where $\delta>0$ is the constant in Assumption 3.2.
Lyapunov's  central limit theorem states that
\begin{align}
\vert n^{-\frac{1}{2}}U^{T}\varepsilon\vert=\vert n^{-\frac{1}{2}}\sum\limits_{i=1}^{n}u_{i}\varepsilon_{i}\vert=O_{p}(1) \notag.
\end{align}
And
\begin{align}
P(J_{3}>\sqrt{logp})=P(\vert n^{-\frac{1}{2}}U^{T}\varepsilon \vert >\hat{\sigma}_{e}\sqrt{logp}) \notag.
\end{align}
According to (1) in \eqref{A.7}: $\sigma_{*}/\sqrt{2} \leq \hat{\sigma}_{e}$, we have
\begin{align}
P(\vert n^{-\frac{1}{2}}U^{T}\varepsilon\vert >\hat{\sigma}_{e}\sqrt{logp})\leq P(\vert n^{-\frac{1}{2}}U^{T}\varepsilon\vert >\sigma_{*}\sqrt{logp/2}) = o_{p}(1) \notag.
\end{align}
Therefore,
\begin{align}
P(J_{3}>\sqrt{logp})=o_{p}(1) \tag{1.16} \label{A.10}.
\end{align}

Due to the sub-Gaussian property of $u_{i}$ and $\sigma_{*}=\sqrt{(\beta_{*}-\beta_{0})^{2}\sigma_{u}^{2}+\sigma_{\varepsilon}^{2}}$, as well as the fact that  $E\vert u_{i}^{2}(\beta_{*}-\beta_{0})/\sigma_{*}\vert ^{3}$ is bounded by a constant, then Lyapunov's central limit theorem implies that \begin{align}
&\sqrt{n}\vert n^{-1}U^{T}U(\beta_{*}-\beta_{0})-\sigma_{u}^{2}(\beta_{*}-\beta_{0})\vert/\sigma_{*} \notag \\
=&\vert n^{-\frac{1}{2}}\sum\limits_{i=1}^{n}[u_{i}^{2}(\beta_{*}-\beta_{0})/\sigma_{*}-E(u_{i}^{2}(\beta_{*}-\beta_{0})/\sigma_{*})]\vert \notag \\
=&O_{p}(1) \notag.
\end{align}
Therefore, according to (1) in \eqref{A.7}: $\hat{\sigma}_{e}\geq \sigma_{*}/\sqrt{2}$, that is $\hat{\sigma}_{e}^{-1}\leq \frac{\sqrt{2}}{\sigma_{*}}$, we have
\begin{align}
&P(\sqrt{n} \vert n^{-1}U^{T}U(\beta_{*}-\beta_{0})-\sigma_{u}^{2}(\beta_{*}-\beta_{0}) \vert \hat{\sigma}_{e}^{-1}>\sqrt{logp}) \notag \\
\leq & P(\sqrt{n} \vert n^{-1}U^{T}U(\beta_{*}-\beta_{0})-\sigma_{u}^{2}(\beta_{*}-\beta_{0}) \vert /\sigma_{*}> \sqrt{logp/2}) \notag \\
=&o_{p}(1) \tag{1.17} \label{A.11}.
\end{align}

To sum up, according to the decomposition of $T_{n}$ and the subadditivity of the probability measure, for any $K>0$, we have
\begin{align}
&P(T_{n}>(K-4)\sqrt{logp}) \notag \\
\geq& P(J_{1}>(K-1)\sqrt{logp})-P(J_{2}>\sqrt{logp})-P(J_{3}>\sqrt{logp})-P(J_{4}>\sqrt{logp}) \tag{*} \label{*}.
\end{align}
According to \eqref{A.8}, \eqref{A.9} and \eqref{A.10}, we have
\begin{align}
\eqref{*}=P(J_{1}>(K-1)\sqrt{logp})-o(1)=P(J_{1}/\sqrt{n}>(K-1)\sqrt{logp/n}) \tag{**} \label{**}.
\end{align}
where,
\begin{align}
J_{1}=&\vert n^{-\frac{1}{2}}U^{T}U(\beta_{*}-\beta_{0})\hat{\sigma}_{e}^{-1}\vert \notag \\
=&\sqrt{n}\vert \sigma_{u}^{2}(\beta_{*}-\beta_{0})\hat{\sigma}_{e}^{-1}+n^{-1}U^{T}U(\beta_{*}-\beta_{0})\hat{\sigma}_{e}^{-1}-\sigma_{u}^{2}(\beta_{*}-\beta_{0})\hat{\sigma}_{e}^{-1}\vert \notag \\
\geq & \sqrt{n}\vert \sigma_{u}^{2}(\beta_{*}-\beta_{0})\hat{\sigma}_{e}^{-1}\vert-\sqrt{n}\vert n^{-1}U^{T}U(\beta_{*}-\beta_{0})\hat{\sigma}_{e}^{-1}-\sigma_{u}^{2}(\beta_{*}-\beta_{0})\hat{\sigma}_{e}^{-1}\vert \notag.
\end{align}
Therefore,
\begin{align}
\eqref{**}
\geq& P(\vert \sigma_{u}^{2}(\beta_{*}-\beta_{0})\vert \hat{\sigma}_{e}^{-1}
>K\sqrt{n^{-1}logp}) \notag \\
-&P(\vert n^{-1}U^{T}U(\beta_{*}-\beta_{0})-\sigma_{u}^{2}(\beta_{*}-\beta_{0})\vert \hat{\sigma}_{e}^{-1} >K\sqrt{n^{-1}logp}) \tag{***} \label{***}.
\end{align}
The  \eqref{A.11} implies that
\begin{align}
\eqref{***}=&P(\vert \sigma_{u}^{2}(\beta_{*}-\beta_{0})\vert \hat{\sigma}_{e}^{-1}
>K\sqrt{n^{-1}logp})-o(1) \notag \\
=&P(\vert \sigma_{u}^{2}(\beta_{*}-\beta_{0})\vert
>K \hat{\sigma}_{e}\sqrt{n^{-1}logp})-o(1) \tag{****} \label{****}.
\end{align}
According to (1) in \eqref{A.7}: $\hat{\sigma}_{e}<2\sigma_{*}$, we get
\begin{align}
\eqref{****}\geq P(\vert \sigma_{u}^{2}(\beta_{*}-\beta_{0})\vert>2K\sigma_{*}\sqrt{n^{-1}logp})-o(1) \notag.
\end{align}
Because $\sigma_{*}=\sqrt{(\beta_{*}-\beta_{0})^{2}\sigma_{u}^{2}+\sigma_{\varepsilon}^{2}}$, there are constants $C_{3}, C_{4}>0$, satisfying that
\begin{align}
\sigma_{*}^{2}\leq C_{3}(\beta_{*}-\beta_{0})^{2}+C_{4} \leq (\sqrt{C_{3}}(\beta_{*}-\beta_{0})+\sqrt{C4})^{2} \notag.
\end{align}
That is $\sigma_{*}\leq \sqrt{C_{3}}\vert \beta_{*}-\beta_{0}\vert+\sqrt{C4}$.
Therefore, the above results show that, for any $K>0$, we have
\begin{align}
&P(T_{n}>(K-4)\sqrt{logp}) \notag \\
\geq &P(\vert \sigma_{u}^{2}(\beta_{*}-\beta_{0})\vert>2K\sigma_{*}\sqrt{n^{-1}logp})-o(1)) \notag \\
\geq &P(\vert \sigma_{u}^{2}(\beta_{*}-\beta_{0})\vert>2K[\sqrt{C_{3}}\vert\beta_{*}-\beta_{0}\vert+\sqrt{C4}]\sqrt{n^{-1}logp})-o(1) \tag{1.18} \label{A.12}.
\end{align}
{\bf{Step 2: derive the behavior of the critical value. }}

Consider the elementary inequalities, for $\xi \sim N(0,1)$, we have
\begin{align}
P(\vert \xi\vert >x)\leq 2 exp\{-x^{2}/2\}, for \  \forall \  x>0 \notag.
\end{align}
In addition,
\begin{align}
P(\vert\xi\vert>x\vert \hat{Q}) \leq 2p\cdot exp[-x^{2}/(2\vert\hat{Q}\vert)] \notag,
\end{align}
where, $\xi\vert \hat{Q}\sim N(0,\hat{Q})$. Therefore,
\begin{align}
\Phi^{-1}(1-\alpha/2;\hat{Q})\leq \sqrt{-2\vert\hat{Q}\vert log(\alpha/2p)} \tag{1.19}. \label{A.13}
\end{align}

Let $C_{5}>0$ be a constant such that $E(u_{1}^{2})\leq C_{5}$.
\begin{align}
P(\vert\hat{Q}\vert\leq 2C_{5})=&P(\vert\hat{Q}-Eu_{1}^{2}+Eu_{1}^{2}\vert\leq 2C_{5}) \notag \\
\geq &P([\vert\hat{Q}-Eu_{1}^{2}\vert+\vert Eu_{1}^{2}\vert]\leq 2C_{5}) \notag \\
\geq & \frac{1}{2}P(\vert\hat{Q}-Eu_{1}^{2}\vert\leq C_{5})+ \frac{1}{2}P(\vert Eu_{1}^{2}\vert\leq C_{5}) \notag.
\end{align}
According to Lemma \ref{lemma 4.3}, we have
\begin{align}
\frac{1}{2}P(\vert\hat{Q}-Eu_{1}^{2}\vert\leq C_{5}) + \frac{1}{2}P(\vert Eu_{1}^{2}\vert\leq C_{5})
=\frac{1}{2}P(O_{p}(\sqrt{n^{-1}logp}) \leq C_{5})+\frac{1}{2}
\rightarrow& 1 \notag.
\end{align}
That is,
\begin{align}
P(\vert\hat{Q}\vert\leq 2C_{5})\rightarrow 1 \tag{1.20} \label{A.14}.
\end{align}
By \eqref{A.13} and \eqref{A.14}, we get
\begin{align}
\Phi^{-1}(1-\alpha/2; \hat{Q})\leq \sqrt{-2\cdot 2\cdot C_{5} log(\alpha/2p) } = 2\sqrt{-C_{5}log(\alpha/2p)} \notag.
\end{align}
Therefore,
\begin{align}
P(\Phi^{-1}(1-\alpha/2; \hat{Q})> 2\sqrt{-C_{5}log(\alpha/2p)})\rightarrow 0,   \ \ {\forall \alpha \in (0,1)} \tag{1.21}. \label{A.15}
\end{align}
Let $K=4\sqrt{C_{5}\vee 1}+8$, then
\begin{align}
&\lim\limits_{p\rightarrow \infty}\frac{(K-4)\sqrt{logp}}{2\sqrt{-C_{5}log(\alpha/2p)}}
=\lim\limits_{p\rightarrow \infty}\frac{(2\sqrt{C_{5}}+2)\sqrt{logp}}{\sqrt{C_{5}log 2p}}
=\lim\limits_{p\rightarrow \infty}\sqrt{2}+\frac{\sqrt{2}}{\sqrt{C_{5}}}
\geq  \sqrt{2} >1  \notag,  \ \ {\forall \alpha \in (0, 1)}.
\end{align}
That is,
\begin{align}
(K-4)\sqrt{logp}>2\sqrt{-C_{5}log(\alpha/2p)}  \tag{1.22} \label{A.16}.
\end{align}
Therefore,
\begin{align}
&P(T_{n}>\Phi^{-1}(1-\alpha/2;\hat{Q})) \notag \\
\geq &P(T_{n}>2\sqrt{-C_{5}log(\alpha/2p)}) \notag \\
\geq& P(T_{n}>(K-4)\sqrt{logp}) \notag \\
\geq &P(\vert \sigma_{u}^{2}(\beta_{*}-\beta_{0})\vert>2K[\sqrt{C_{3}}\vert\beta_{*}-\beta_{0}\vert+\sqrt{C4}]\sqrt{n^{-1}logp})-o(1) \notag
%\geq & P(\vert e_{j*}^{T}\Sigma_{U}(\gamma-\beta_{G}^{0})\vert >2K[\sqrt{C_{3}}\|(\gamma-\beta_{G}^{0})\|_{2}+\sqrt{C_{4}}]\sqrt{n^{-1}logp}) \notag.
\end{align}
where the first inequality holds by \eqref{A.15}, the second inequality holds by \eqref{A.16}, the third inequality holds by \eqref{A.12}.
Let $K_{1}=2K\sqrt{C_{3}}$, $K_{2}=2K\sqrt{C_{4}}$, then
\begin{align}
&P(T_{n}>\Phi^{-1}(1-\alpha;\hat{Q})) \notag \\
\geq &P(\vert \sigma_{u}^{2}(\beta_{*}-\beta_{0})\vert>[K_{1}\vert\beta_{*}-\beta_{0}\vert+K_{2}]\sqrt{n^{-1}logp})-o(1) \notag.
\end{align}
%According to the assumption  $\|\Sigma_{u}(\gamma-\beta_{G}^{0})\|_{\infty}\geq\sqrt{n^{-1}logp}(K_{1}\|(\gamma-\beta_{G}^{0})\|_{2}+K_{2})$ in Assumption 2.1, we get
Whenever $\vert\sigma_{u}^{2}(\beta_{*}-\beta_{0})\vert\geq \sqrt{n^{-1}logp}(K_{1}\vert\beta_{*}-\beta_{0}\vert+K_{2})$ in Theorem 2, we get
\begin{align}
P(T_{n}>\Phi^{-1}&(1-\alpha/2;\hat{Q}))\rightarrow 1 \notag.
\end{align}

The proof is complete.
\end{proof}
 \bibliographystyle{elsarticle-harv}
\bibliography{sn-bibliography11}


%\begin{thebibliography}{00}


%\end{thebibliography}